\documentclass[twocolumn]{aastex631}
\usepackage{amsmath}
\usepackage{mathtools}
\usepackage{hyperref}

\newcommand{\specnotation}[2]{\ensuremath{\rm #1 \, {\scriptstyle #2}}}
\newcommand{\molH}{\ensuremath{\rm H_2}}

\newcommand{\kms}{km\,s$^{-1}$}
\newcommand{\cmmt}{cm$^{-2}$}

\newcommand{\lten}{{\rm log}_{10}}

\newcommand{\NmolH}{\ensuremath{N_{\rm H_2}}}

\newcommand{\NCO}{\ensuremath{N_{\rm CO}}}

\newcommand{\FUSE}{\emph{FUSE}}
\newcommand{\sknum}[2]{{Sk$\,\,$-{#1}$^\circ${#2}}}
\newcommand{\isoCO}[1]{$^{#1}$CO}
\newcommand{\letCO}[1]{\ensuremath{{#1}_{\rm CO}}}
\newcommand{\isoratio}[3]{$^{#2}${#1}/$^{#3}${#1}}
\newcommand{\dd}{{\rm d}}

\graphicspath{{./}{figures/}}

\shortauthors{Tchernyshyov, Werk, \& Roman-Duval}

\begin{document}

\title{An Ultraviolet Study of CO Chemistry in the Magellanic Clouds}

\author[0000-0003-0789-9939]{Kirill Tchernyshyov}
\affiliation{Department of Astronomy, University of Washington, Seattle, WA, USA}
\correspondingauthor{Kirill Tchernyshyov}
\email{ktcherny@gmail.com}

\author[0000-0002-0355-0134]{Jessica K. Werk}
\affiliation{Department of Astronomy, University of Washington, Seattle, WA, USA}

\author[0000-0001-6326-7069]{Julia Roman-Duval}
\affiliation{Space Telescope Science Institute, Baltimore, MD, USA}

\begin{abstract}
How does molecular cloud chemistry change with metallicity?
In this work, we study the relation between molecular hydrogen (\molH) and carbon monoxide (CO) at $1/2$ and $1/5$ solar metallicity using ultraviolet absorption spectroscopy obtained as part of the UV Legacy Library of Young Stars as Essential Standards (ULLYSES) Hubble Space Telescope (HST) program.
We determine CO column densities or upper limits for a sample of  50 lines of sight through the Large and Small Magellanic Clouds (LMC and SMC). 
\isoCO{12}\ is detected along eight lines of sight and \isoCO{13}\ is detected along two. 
Combining our new CO column densities with \NmolH\ measurements from the literature, we find that the evolution of $\NCO(\NmolH)$ from the Milky Way to the LMC and SMC is a relatively shallow function of metallicity. 
Taking $\NCO>3\times10^{15}$ \cmmt\ as a threshold value above which CO emission is likely to be detectable at the distance of the Magellanic Clouds, the $\lten \NmolH$\ at which a sightline has a 50\% probability of having $\NCO$\ above this threshold is 20.8 in the Milky Way, 20.9 in the LMC, and 21.1 in the SMC. 
This is an 0.3 dex change in threshold $\lten \NmolH$\ over an 0.7 dex change in metallicity. 
We compare our measurements with $\NCO(\NmolH)$\ relations from literature chemical models and find that the measured relations agree best with models in which the dynamical timescale is longer than the chemical timescale for \molH\ but shorter than the chemical timescale for CO. 
\end{abstract}

\section{Introduction}
\label{sec:intro}

There is a need for simultaneous observations of molecular hydrogen (\molH) and carbon monoxide (CO) in gas with sub-solar metallicity. 
CO emission is the most heavily used \molH\ mass tracer, and the conversion factor (\letCO{X} or \letCO{\alpha}) between these two quantities  has been studied a great deal using different combinations of other emitting gas mass tracers.
Recent works using this approach have found the slope of the correlation between the log of metallicity and the log of \letCO{X}\ to be -3.4 \citep{Madden:2020aa-XCO}, -1.6 \citep{Accurso:2017aa-XCO}, or variable from galaxy to galaxy within the range spanned between the two previous references \citep{Ramambason:2024-XCO-split}.
Theoretical calculations of the slope have not converged either.
\citet{Thompson:2024-neq-CO-H2} predict -2.0, \citet{Gong:2020-TIGRESS-XCO-variation} predict -0.7, and \citet{Hu:2023-lowZ-dust-evo} do not make an explicit prediction but produce a simulated 0.1 $Z_\odot$\ galaxy that would imply a slope of -0.2. 
Taking the observational and theoretical work together, we can conclude that $\letCO{X}$ in a $0.1 Z_{\odot}$ galaxy is somewhere between 1.4 and 2500 times greater than $\letCO{X}$ in the Milky Way.

A problem common to all of the above approaches is an excess of independent variables.
Emission-based observational studies require independent estimates of \molH\ mass at the location of CO emission. 
If these estimates are based on tracers such as dust emission, they then themselves depend on uncertain metallicity-dependent calibration factors (e.g., \citealt{Roman-Duval:2014aa,Roman-Duval:2017aa,Roman-Duval:2022aa,Clark:2023-DGR-Z-Sigma-variation}).
It is possible to derive \letCO{X}\ given an assumed dust-emission-to-gas ratio or vice versa, but attempts to solve for both run into the problem of collinearity. 
The uncertainty in theoretical predictions for \letCO{X}\ is downstream of uncertainty on physical conditions in molecular gas.
For example, the combined effect of the cosmic ray ionization rate and the dwell time of a Lagrangian gas parcel in a molecular cloud is not monotonic: increasing the ionization rate reduces the CO-to-\molH\ abundance ratio for dwell times longer than the \molH\ chemical timescale, but increases it for dwell times between the CO and \molH\ timescales \citep{Bialy:2015,Bisbas:2015,Bisbas:2017,Hu:2021uj}.

To solve this specific problem, we need simultaneous observations of CO and \molH\ with no dependence on uncertain calibration factors. 
This can be done using ultraviolet (UV) absorption spectroscopy, which can provide CO and \molH\ column densities along pencil beam lines of sight through gas between an observer and a background source.
If the goal is to determine $\letCO{X}(Z)$, column density measurements have their own problems: the system geometry differs from that of emission measurements (pencil beams vs. telescope beams) and non-trivial excitation analysis and radiative transfer are required to go from CO column densities to emission intensities.
Nevertheless, CO and \molH\ column density measurements on their own give a rough sense of how \letCO{X}\ scales with metallicity and are an absolute calibration point for theoretical models, which can then be used to produce better-informed predictions for \letCO{X}. 

While absorption measurements of CO and \molH\ are powerful, they are also hard to get.
Individual CO absorption lines are intrinsically narrow, are clustered in wavelength into ``bands'', and except at very high \molH\ column densities tend to be weak.
In spectra with resolution less than $R\sim 100,000$, a CO band consisting of multiple narrow lines will appear as a single, typically shallow, feature with total width about 100 \kms.
Observing \molH\ absorption at redshift zero requires coverage of the Lyman UV (912 to 1215 \AA), a range which has not been (efficiently) observable since the decommissioning of the Far Ultraviolet Spectroscopic Explorer (\FUSE) in 2007. 
Lines of sight with high \molH\ column densities will also contain a large column of dust, meaning that background sources will need to be exceptionally UV-bright in order to be usable for spectroscopy. 
These specific challenges can be circumvented by working at higher redshift, but observing CO absorption at high redshift creates a new challenge: the need for a lucky combination of a bright quasar behind a (typically small) dense molecular cloud. 

Almost all studies of \molH\ and CO absorption have been in the Milky Way, where they serve as an anchor for models of molecular cloud chemistry at Solar metallicity \citep{Rachford:2002wo,Crenny:2004ul,Burgh:2007vc,Sonnentrucker:2007ux,Sheffer:2008vc,Burgh:2010wd}.
There have been a few studies of CO absorption in the sub-solar metallicity Magellanic Clouds \citep{Bluhm:2001aa-CO-Sk-69-246,Koenigsberger:2001-implausible-CO-detection,Andre:2004aa} and in low metallicity damped Lyman-$\alpha$ systems \citep{Balashev:2017-CO-dark-DLA,Ranjan:2018-strong-H2-absorber}. 
These studies have focused on one to a few lines of sight at a time.
Even taken together, these studies do not provide a large enough sample to reveal how the correlation between \molH\ and CO depends on metallicity.

In this work, we undertake a systematic study of \molH\ and CO absorption in a sub-solar metallicity environment.
We measure CO column densities or upper limits along 50 lines of sight through the Large and Small Magellanic Clouds, which have metallicities 1/2 and 1/5th solar, respectively \citep{Russell:1992aa}. 
The spectra come from the UV Legacy Library of Young Stars as Essential Standards (ULLYSES) program \citep{Roman-Duval:2020aa-ULLYSES-citation}.
By combining these measurements with archival \molH\ column densities along the same sightlines, we produce a powerful new constraint for models of molecular cloud chemistry and the dependence of $\letCO{X}$ on metallicity.
We describe the sample of ultraviolet spectra in \S\ref{sec:data}, measure CO column densities from the spectra in \S\ref{sec:analysis}, analyze the correlation between \molH\ and CO column densities in \S\ref{sec:results}, discuss limitations and possible consequences of our analysis in \S\ref{sec:discussion}, and summarize our results in \S\ref{sec:conclusion}.

\section{Data} \label{sec:data}

\subsection{Sample selection} \label{sec:data:sample}

Our analysis requires direct measurements of \molH\ and CO column densities.
\citet{Welty:2012vl} compiles \molH\ column density measurements based on absorption features in \FUSE\ spectra towards stars in the LMC and SMC. 
Our sample consists of every star in the LMC and SMC that (1) is known to have a \molH\ column density greater than $10^{19}$ \cmmt\ and (2) has an archival spectrum from ULLYSES or other programs covering 1400 to 1500 \AA\ with at least medium resolution ($R\gtrsim15,000$).
This selection yields 26 stars in the LMC and 24 stars in the SMC.
Basic information about the targets is listed in \autoref{tab:sample-def}.

\startlongtable
\begin{deluxetable*}{lllcccc}
\tablecaption{Target properties \label{tab:sample-def}}
\tablehead{
Simbad ID & ULLYSES ID & \colhead{Galaxy} & \colhead{R.A.} & \colhead{Declination} &  \colhead{Gratings} &  \colhead{Log$_{10}$\NmolH}\\
& & & (deg) & (deg) & & [\cmmt]}
\startdata
SK -71 8 & SK-71D8 & LMC & 76.84691 & -71.19832 & STIS E140M & 19.02\\
SK -69 220 & SK-69D220 & LMC & 84.18206 & -69.49651 & STIS E140M & 19.04\\
SK -67 105 & SK-67D105 & LMC & 81.52579 & -67.18244 & STIS E140M & 19.13\\
PGMW 3070 & PGMW-3070 & LMC & 74.18032 & -66.41736 & STIS E140M & 19.21\\
SK -66 35 & SK-66D35 & LMC & 74.2685 & -66.57738 & STIS E140M & 19.3\\
SK -67 5 & SK-67D5 & LMC & 72.57886 & -67.66057 & STIS E140M, E140H & 19.46\\
SK -68 52 & SK-68D52 & LMC & 76.83509 & -68.53571 & STIS E140M & 19.47\\
BI 184 & BI-184 & LMC & 82.62776 & -71.04211 & COS & 19.65\\
SK -69 246 & SK-69D246 & LMC & 84.72241 & -69.03358 & STIS E140M & 19.71\\
VFTS 72 & VFTS-72 & LMC & 84.39358 & -69.01949 & COS & 19.76\\
SK -68 135 & SK-68D135 & LMC & 84.45473 & -68.91712 & STIS E140M & 19.87\\
SK -70 115 & SK-70D115 & LMC & 87.20688 & -70.06607 & STIS E140M, E140H & 19.94\\
SK -68 155 & SK-68D155 & LMC & 85.72893 & -68.94847 & COS & 19.99\\
BI 237 & BI-237 & LMC & 84.06097 & -67.65532 & COS & 20.05\\
SK -69 224 & SK-69D224 & LMC & 84.22987 & -69.19379 & COS & 20.07\\
SK -68 73 & SK-68D73 & LMC & 80.74911 & -68.02962 & COS, STIS E140H & 20.09\\
SK -68 140 & SK-68D140 & LMC & 84.73821 & -68.9481 & COS & 20.11\\
SK -71 50 & SK-71D50 & LMC & 85.17989 & -71.48352 & STIS E140M & 20.13\\
SK -68 129 & SK-68D129 & LMC & 84.11156 & -68.9589 & COS & 20.2\\
SK -66 19 & SK-66D19 & LMC & 73.97479 & -66.41649 & COS & 20.2\\
SK -71 46 & SK-71D46 & LMC & 82.95662 & -71.06057 & COS & 20.21\\
SK -70 79 & SK-70D79 & LMC & 76.65531 & -70.49005 & STIS E140M & 20.26\\
SK -69 279 & SK-69D279 & LMC & 85.43606 & -69.58747 & COS & 20.31\\
SK-68 137 & SK-68D137 & LMC & 84.60283 & -68.87582 & COS & 20.37\\
SK -68 26 & SK-68D26 & LMC & 75.38436 & -68.17859 & COS & 20.38\\
SK -67 2 & SK-67D2 & LMC & 71.76854 & -67.11476 & STIS E140M & 20.95\\
AV 6 & AV-6 & SMC & 11.32583 & -73.25654 & COS & 19.11\\
AV 479 & AV-479 & SMC & 18.70942 & -73.33829 & COS, STIS E140H & 19.15\\
AV 104 & AV-104 & SMC & 12.91012 & -72.80169 & STIS E140M & 19.23\\
AV 261 & AV-261 & SMC & 15.24466 & -72.51395 & COS & 19.26\\
AV 210 & AV-210 & SMC & 14.6491 & -72.27362 & STIS E140M & 19.36\\
AV 388 & AV-388 & SMC & 16.41471 & -72.49082 & COS, STIS E140M & 19.4\\
AV 207 & AV-207 & SMC & 14.63829 & -71.92964 & COS & 19.4\\
AV 95 & AV-95 & SMC & 12.84 & -72.73747 & STIS E140M & 19.4\\
2dFS 999 & 2DFS-999 & SMC & 13.63398 & -72.74324 & COS & 19.41\\
AV 304 & AV-304 & SMC & 15.58951 & -72.65405 & COS & 19.57\\
AV 215 & AV-215 & SMC & 14.73177 & -72.53569 & STIS E140M & 19.59\\
AV 170 & AV-170 & SMC & 13.92679 & -73.29181 & STIS E140M & 19.67\\
AV 490 & AV-490 & SMC & 19.27144 & -73.44334 & STIS E140M & 19.78\\
AV 175 & AV-175 & SMC & 14.15866 & -72.6098 & COS & 19.8\\
AV 435 & AV-435 & SMC & 17.07462 & -71.99843 & COS & 19.88\\
AV 80 & AV-80 & SMC & 12.68256 & -72.79487 & STIS E140M & 20.08\\
AV 16 & AV-16 & SMC & 11.72929 & -73.14282 & STIS E140M & 20.22\\
AV 472 & AV-472 & SMC & 18.25781 & -72.76346 & COS & 20.3\\
AV 18 & AV-18 & SMC & 11.8009 & -73.1092 & STIS E140M, E140H & 20.36\\
Cl* NGC 346 ELS 26 & NGC346-ELS-026 & SMC & 14.55869 & -72.17897 & STIS E140M & 20.42\\
AV 26 & AV-26 & SMC & 11.95853 & -73.13918 & STIS E140M, E140H & 20.63\\
SK 191 & SK-191 & SMC & 25.42528 & -73.84394 & COS, STIS E140M & 20.65\\
AV 456 & AV-456 & SMC & 17.73232 & -72.71562 & COS, STIS E140H & 20.93\\
AV 476 & AV-476 & SMC & 18.42687 & -73.29153 & COS, STIS E140M & 20.95\\
\enddata
\tablecomments{\molH\ column densities are from \citet{Welty:2012vl}. \NmolH\ uncertainties are not given for individual lines of sight but are stated to be 10-20\%, corresponding to logarithmic uncertainties of 0.04 to 0.08. In the \emph{Gratings} column, ``COS'' indicates the availability of data taken with both the G130M and G160M gratings.}
\end{deluxetable*}

All stars in the sample have observations taken with the STIS and/or COS spectrographs on the \emph{Hubble Space Telescope} \citep{Kimble:1998_STIS_performance,Woodgate:1998_STIS_design,Green_2012_COS_citation} and uniformly processed and released by the ULLYSES project (\dataset[10.17909/t9-jzeh-xy14]{https://doi.org/10.17909/t9-jzeh-xy14}, \citealt{Roman-Duval:2020aa-ULLYSES-citation}).
Some of the targets were originally observed for unrelated programs while others were observed specifically for ULLYSES.
We use the single instrument and grating coadded spectra (\texttt{cspec.fits} files) from the current most up-to-date release, DR7.

The COS spectra were recorded with the G130M and G160M gratings, which provide resolution $R\approx 16,000$.
STIS spectra were mostly recorded using the E140M grating, which provides resolution $R\approx 36,000$.
These settings provide coverage of the CO A-X bands \citep{Morton:1994ty}.
Seven targets also have observations with the STIS E140H grating, which provides resolution $R\approx 100,000$ but covers a narrower wavelength range per setting.

CO transition wavelengths, oscillator strengths, and damping constants are taken from \citet{Morton:1994ty}, \citet{Eidelsberg:2003-intersystem-CO-transitions}, and \citet{Dapra:2016:CO}.

\section{Analysis} \label{sec:analysis}

We measure \isoCO{12} and, in some cases, \isoCO{13} column densities using Voigt profile fitting.
The analysis is done in two parts: (1) building an informative model for the shape of the stellar continuum and (2) the actual Voigt profile fitting.
The continuum model is probabilistic rather than fixed. 
The shape of the continuum at the location of CO absorption is marginalized over during the profile fitting step using the continuum model derived in the first step as a prior.

An important pair of simplifying assumptions in our analysis is (1) that there is at most one detectable CO component along each sightline and (2) that this component will be found at LMC or SMC velocities. 
Put another way, we assume that sightlines pass through at most one CO-containing molecular cloud in the MCs and that the Milky Way foreground contains no CO.
At Milky Way velocities, this assumption is justified by high latitude \molH\ column densities being below values at which CO is typically seen \citep{Gillmon:2006tl,Sheffer:2008vc}.

\citet{Welty:2006MC-molecules} search for CH absorption in high resolution optical spectra toward 20 LMC and SMC targets. 
In 16 of 20 cases, they find at most one velocity component with detectable CH; this includes AV-456. 
The remaining targets have two components, one of which dominates the total column along the line of sight.
Toward \sknum{67}{2}, for example, the column density of the strong component is about 20 times greater than that of the weak component.
Visual inspection of the spectra yields no evidence for CO in the Milky Way foreground or for more than one component of LMC or SMC CO along a sightline.
We acknowledge that the lack of CO absorption systems with clear velocity sub-structure may reflect the limited spectral resolution of the available spectra compared to the typical width of individual CO absorption components.

\subsection{Continuum modeling} \label{sec:analysis:continuum}

In this step of the analysis, we create a physically-informed, probabilistic model for the shape of a target's stellar continuum.
This model combines a theoretical stellar atmosphere spectrum with polynomial and Gaussian process correction terms.
We first fit the theoretical stellar atmosphere spectra to parts of the target's observed spectrum that are free of CO and other ISM absorption.
Several regions in the spectrum that are free of ISM absorption are not included in the atmosphere fit. 
These holdout regions are then used to learn a Gaussian process model for the structure of residuals between the best fit theoretical spectrum and the observed spectrum.

The functional form of the model flux $\bf{f}_\lambda$ in the atmosphere fit is
\begin{equation}
    {\bf f}_\lambda(v_0, v\sin i, {\bf a}, {\bf b}) = {\bf M}(v_0, v\sin i) {\bf a} + b_2 \lambda^2 + b_1 \lambda + b_0.
\end{equation}
${\bf M}$ is a matrix of candidate stellar atmosphere model spectra that have been shifted by a velocity $v_0$ and convolved with a rotational broadening kernel with projected rotation speed $v \sin i$. 
We use LMC and SMC metallicity OB and Wolf-Rayet star spectra calculated from the PoWR atmosphere models\footnote{Available at \href{https://www.astro.physik.uni-potsdam.de/~wrh/PoWR/powrgrid1.php}{https://www.astro.physik.uni-potsdam.de/~wrh/PoWR/powrgrid1.php}.} \citep{Sander_2012_PoWR_WC,Todt_2015_PoWR_WN,Hainich_2019_PoWR_OB}. 
We require the coefficients $\bf{a}$ to be non-negative while the polynomial coefficients $b_1$, $b_2$, and $b_3$ can be arbitrary.
In practice, the non-negativity constraint leads to $\bf{a}$ having a very small number of non-zero entries, typically one or two.

We solve for the best-fit model spectrum in two stages: brute force grid search over a physically motivated range in $v_0$ and $v \sin i$ followed by non-linear optimization from the best-fitting grid point. 
For fixed $v_0$ and $v \sin i$, the flux model is linear in the parameters ${\bf a}$ and $b_1$, $b_2$, and $b_3$. 
As a result, in both the grid search and non-linear optimization steps we can use constrained linear least squares to find the optimal linear parameters for a given pair of $v_0$ and $v \sin i$.
We use \texttt{scipy.optimize.least\_squares} as the linear solver and \texttt{scipy.optimize.fmin\_l\_bfgs\_b} as the non-linear solver for the second step. 
The model spectrum produced by this procedure is our initial estimate of the stellar continuum. 

\begin{figure*}
\centering
\includegraphics{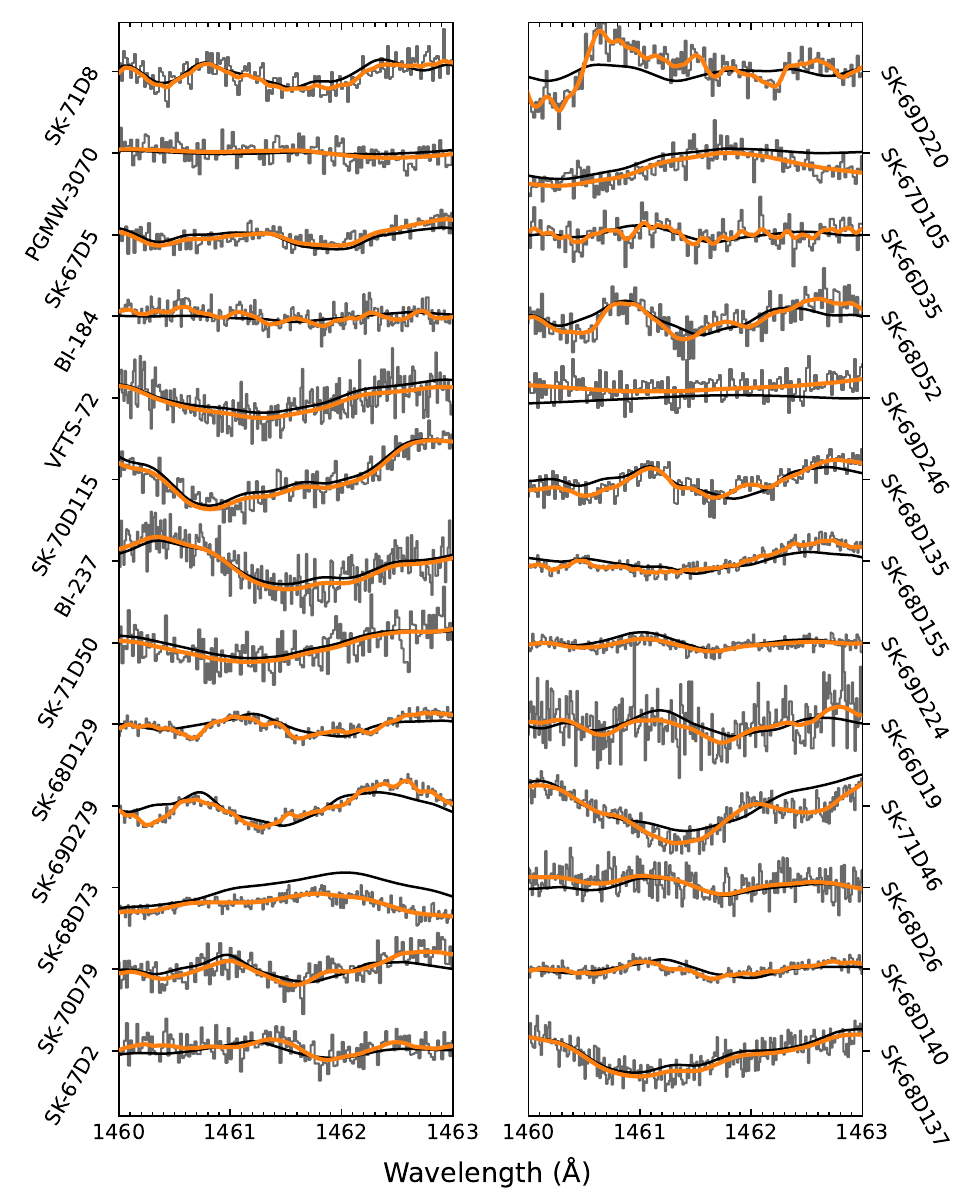}
\caption{Stellar atmosphere model (smooth black line) and stellar atmosphere plus Gaussian process (orange line) continuum fits to spectra of all LMC targets (stepped black line) in an ISM absorption-free holdout region. Spectra are normalized so that the median of the stellar atmosphere model within the visible range is 1 and are offset from each other by 0.5. Zero flux for a given star is two stars down; for example, zero flux for \sknum{68}{73} is at the tick marked \sknum{67}{2}. The part of the spectrum shown here is excluded from the stellar atmosphere model fit but is used for estimating the amplitude and length scale of the Gaussian process. The difference between the orange and black lines can be thought of as a correction to the stellar atmosphere model fits. The size and typical width of ``bumps'' in the corrections varies considerably from star to star.}
\label{fig:contfit-lmc-holdout}
\end{figure*}

\begin{figure*}
\centering
\includegraphics{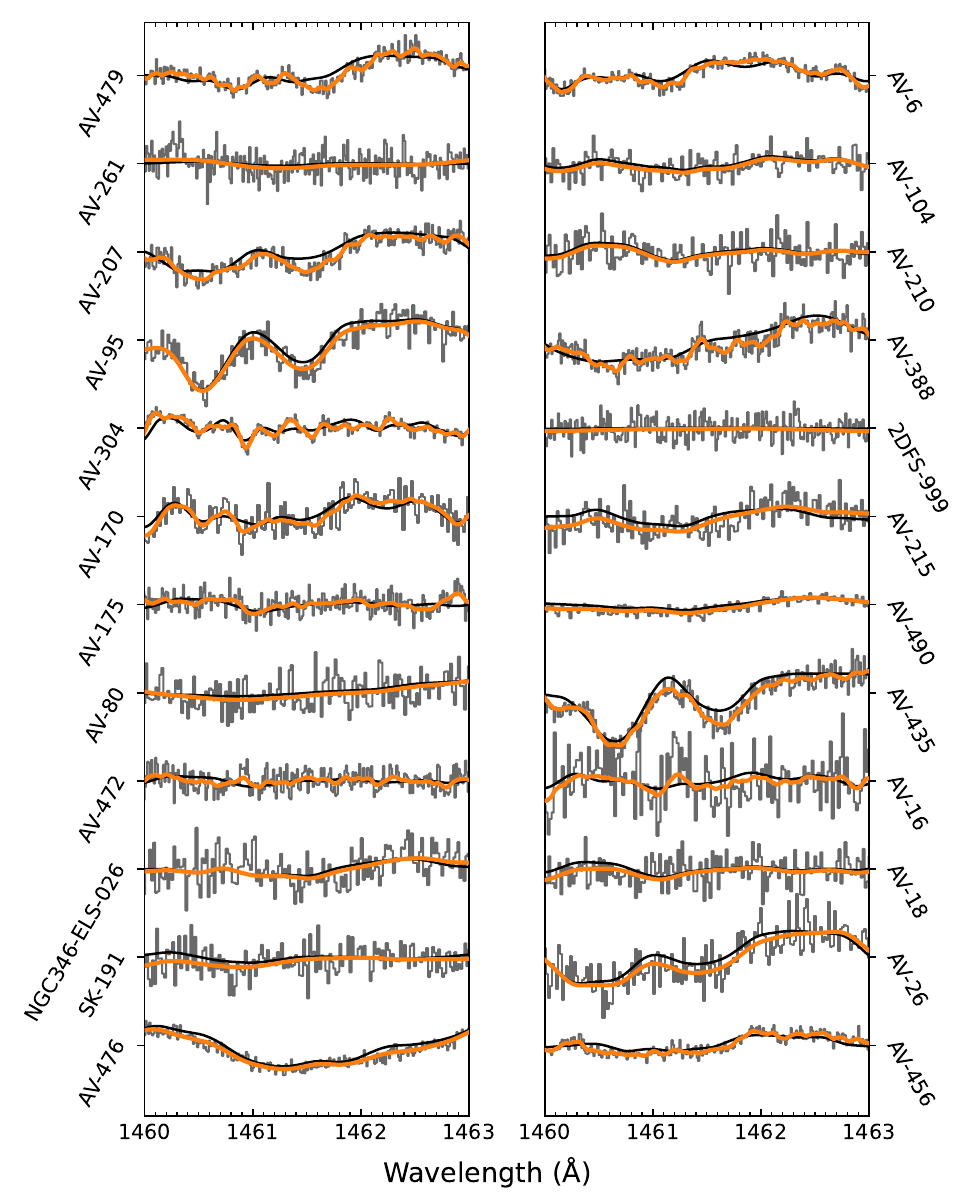}
\caption{Stellar atmosphere model (smooth black line) and stellar atmosphere plus Gaussian process (orange line) continuum fits to spectra of all SMC targets (stepped black line) in an ISM absorption-free holdout region. See \autoref{fig:contfit-lmc-holdout} for a description of figure elements. }
\label{fig:contfit-smc-holdout}
\end{figure*}

Figures \ref{fig:contfit-lmc-holdout} and \ref{fig:contfit-smc-holdout} show stellar atmosphere model fits (black lines) for LMC and SMC stars in one of the ISM-absorption-free holdout regions. 
For most of the stars, the models capture the general shape of the stellar pseudocontinuum. 
For the best cases, such as \sknum{70}{115}\ in the LMC and AV95 in the SMC, the difference between the models and observations is small in amplitude and broad in wavelength. 
For other stars, such as \sknum{68}{129} in the LMC and AV304 in the SMC, the models have too few or too many stellar absorption features or have subtly incorrect absorption line shapes. 
These discrepancies have similar widths to CO bands.
Attempts to correct the continuum fit by masking CO absorption and fitting a polynomial to the surrounding CO-free parts of the spectrum could therefore completely miss one of these un-modeled stellar absorption features.
To avoid this pitfall, we will fit the correction to the stellar atmosphere model together with the CO absorption.

In all cases, the residuals between the stellar atmosphere models and observations are structured. 
Colloquially, they can be described as consisting of ``bumps''.
The typical width (\emph{length scale}) and height (\emph{amplitude}) of the bumps are fairly consistent for each star. 
By learning a star's length scales and amplitudes, we can create a prior probability distribution over the difference between the true stellar pseudocontinuum and our best-fit stellar atmosphere models. 

Let $\bf{y}$ be the observed flux for a star, let $\bf{f}$ be the best-fit stellar atmosphere plus polynomial model for the star, and let $\bf{r}$ be the correction from $\bf{f}$ to the actual stellar pseudocontinuum. 
In areas with no ISM absorption, $\bf{y}-\bf{f}$ is a noisy estimate of $\bf{r}$.
By learning the structure of $\bf{r}$ in areas with no ISM absorption, we can better inform estimates of $\bf{r}$ in areas that do contain ISM absorption.
We characterize the structure of $\bf{r}$ by its covariance matrix $\bf{K}$.
The prior probability distribution over $\bf{r}$ is a multivariate Gaussian distribution with this covariance matrix and mean zero. 
This type of probability distribution over functions is known as a Gaussian process.

We assume that the covariance of the residuals is stationary, meaning that the covariance between points with wavelengths $\lambda_1$ and $\lambda_2$ is just a function of $\vert \lambda_1 - \lambda_2 \vert$.
The covariance function we use is the sum of two \emph{Matern $3/2$ kernels}. 
The Matern $3/2$ kernel has the functional form 
\begin{equation}
    K(d=\lambda'-\lambda) = \sigma^2\left(1+\frac{\sqrt{3}d}{\ell}\right) \exp\left[-\frac{\sqrt{3}d}{\ell} \right],
\end{equation}
where $\ell$ is a length scale parameter and $\sigma$ is an amplitude parameter.
The first kernel is given a length scale of 10 \AA\ and is meant to capture broad differences between the initial model and observed spectrum.

We determine the parameters of the second kernel by fitting the residuals in the ISM absorption-free holdout regions.
In this case, ``fitting'' means using non-linear optimization to find the values of the second kernel's amplitude and length scale parameters that maximize the Gaussian process likelihood function. 
We use the \texttt{tinygp} Gaussian process implementation and \texttt{scipy.optimize.fmin\_l\_bfgs\_b} as the non-linear optimizer.

The orange lines in Figures \ref{fig:contfit-lmc-holdout} and \ref{fig:contfit-smc-holdout} are stellar atmosphere plus Gaussian process continuum models for the LMC and SMC targets.
The Gaussian process parameters for these continuum models are optimized for each star using the procedure described above.
Stars whose stellar atmosphere model residuals are more ``wiggly'' have shorter length scales and larger amplitudes than stars whose residuals are more broad and flat.

\begin{figure*}
\centering
\includegraphics{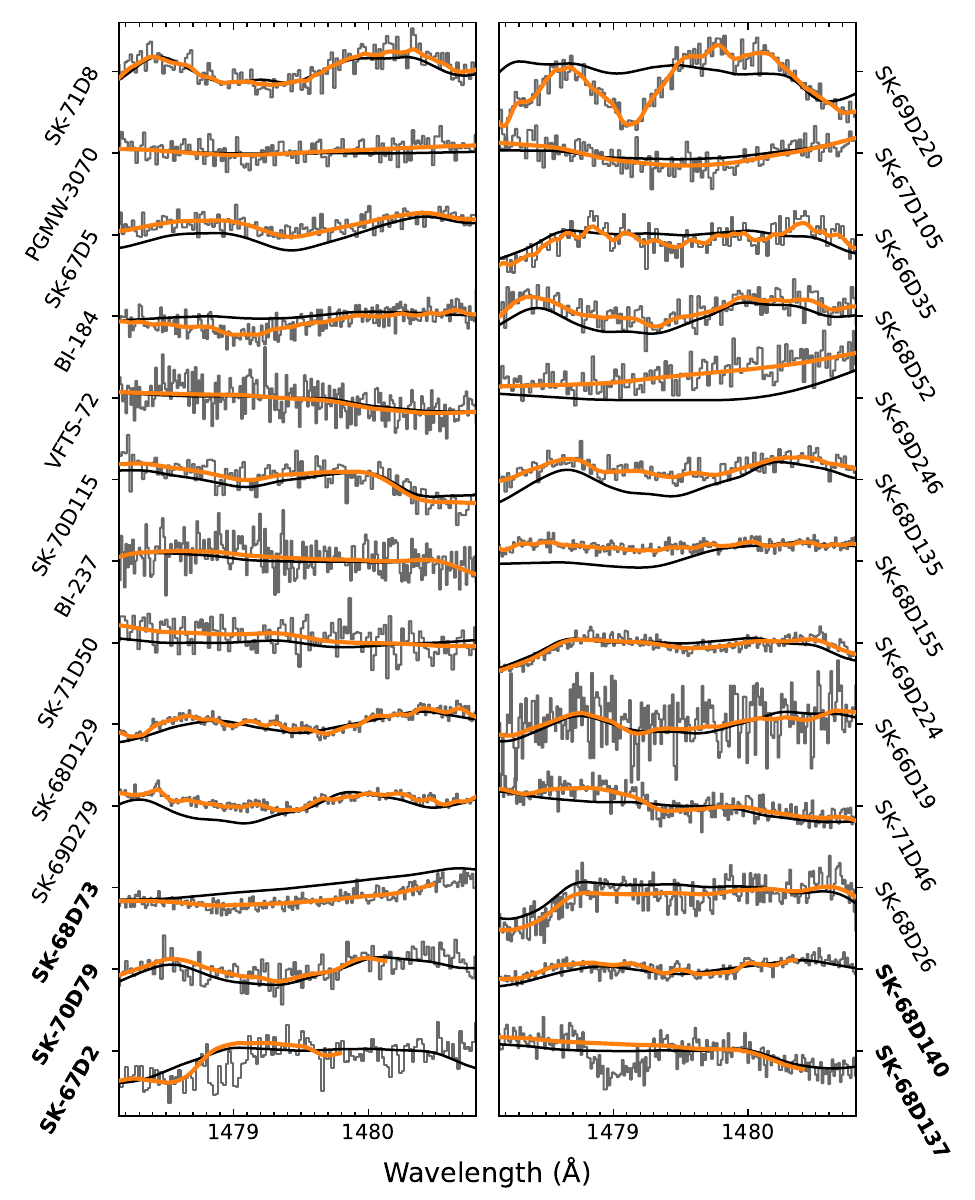}
\caption{Stellar atmosphere model (smooth black line) and stellar atmosphere plus Gaussian process (orange line) continuum fits to spectra of all LMC targets (stepped black line) at the position of the CO A-X(2-0) band. The names of targets with a CO detection (based on all CO bands, not just this one) are bolded. The part of the spectrum shown here is excluded from the stellar atmosphere model fit. For detections, CO absorption and Gaussian process corrections to the initial stellar atmosphere model are fit simultaneously. The absorption fits do not use the entire spectral range shown here. This is why the full continuum fits for targets with detected CO do not extend all the way across their panels.}
\label{fig:contfit-lmc-co}
\end{figure*}

\begin{figure*}
\centering
\includegraphics{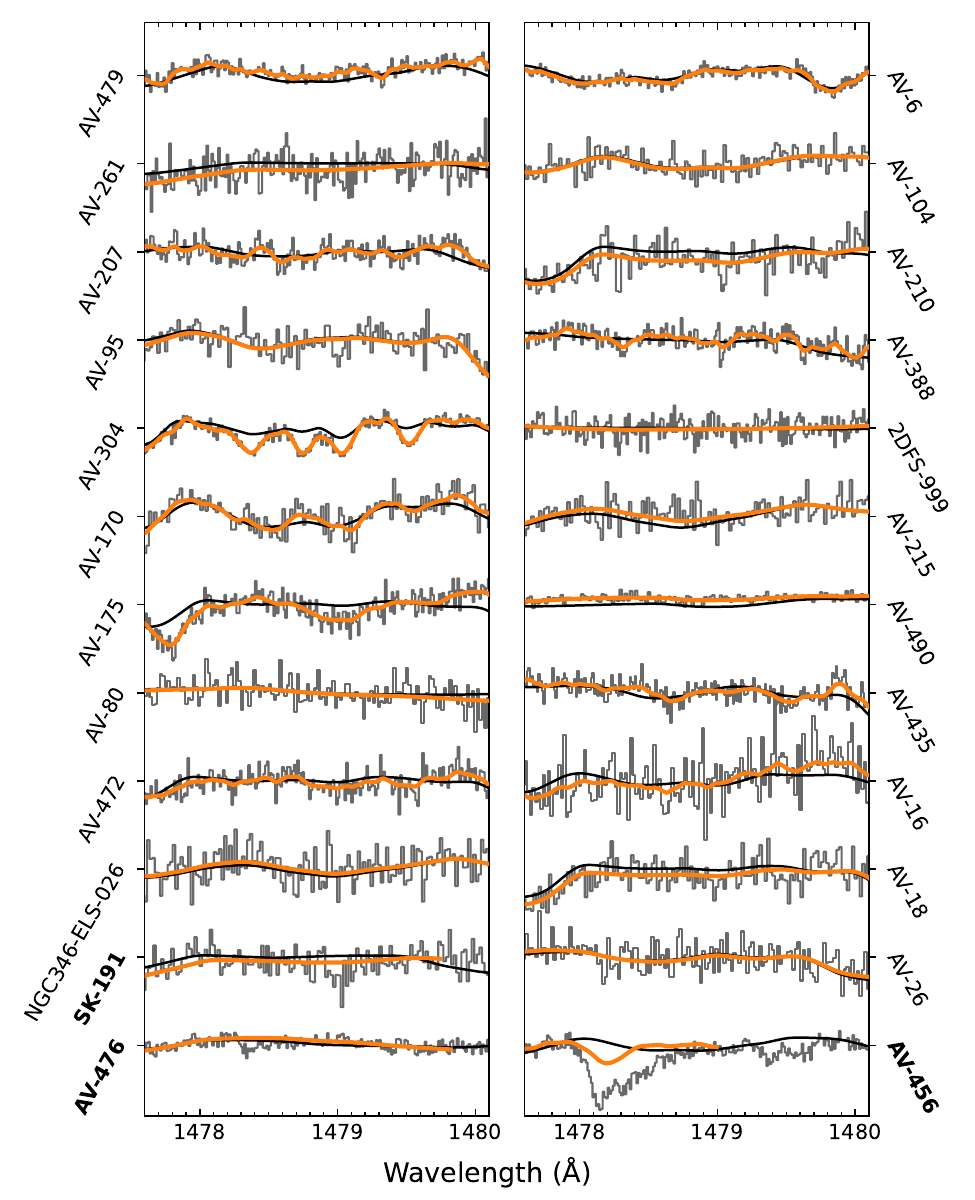}
\caption{Stellar atmosphere model (smooth black line) and stellar atmosphere plus Gaussian process (orange line) continuum fits to spectra of all SMC targets (stepped black line) at the position of the CO A-X(2-0) band. See \autoref{fig:contfit-lmc-co} for a description of figure elements.}
\label{fig:contfit-smc-co}
\end{figure*}

In regions without ISM absorption, the continuum correction estimate $\bf{r}$ is essentially just a smoothed version of the residuals $\bf{y} - \bf{f}$ with a smoothing scale that is controlled by the kernel parameters.
When ISM absorption is present, the residual are no longer a noisy estimate of the continuum correction. 
However, a Gaussian process prior for $\bf{r}$ can still be used so long as we have a model for the ISM absorption.
This is possible because multiplying $\bf{r}$ by a model transmittance vector is a linear transformation and linear transformations applied to a Gaussian process yield a different Gaussian process.
Let $\bf{a}$ be a model transmittance vector and let $\bf{A}$ be the diagonal matrix $\text{diag}(\bf{a})$.
If $\bf{K}$ is the covariance matrix for $\bf{r}$ without absorption, then the covariance matrix for $\bf{r}$ with absorption is $\bf{A K A}$. 
The math involved in this modification of the Gaussian process is explained further in \citet{Tchernyshyov:2020-amlc}.

Figures \ref{fig:contfit-lmc-co} and \ref{fig:contfit-smc-co} show stellar atmosphere and full continuum model fits around the CO A-X(2-0) band. 
Note that the full fits for targets with CO detections are derived using a simultaneous continuum and absorption model fit to multiple CO absorption bands; the absorption model is described in the next section.
The Gaussian process prior inherits the ``bumpiness'' properties derived in the holdout regions
The actual shape of the full continuum model is informed by local information in the spectrum and by the need for a single set of CO parameters to fit multiple CO bands. 

Targets such as AV304 and AV175 in the SMC demonstrate why local information is necessary to analyze this spectral region for this sample of OB stars. 
If we were to mask out the region that could contain CO absorption for an SMC target, 1478 to 1479 \AA, and then fit a polynomial to either side of that region, the fit would not include the complex stellar features found between these wavelengths.
The physically informed stellar atmosphere model also fails to correctly reproduce the features.
A simultaneous continuum and absorption fit is \emph{necessary} to analyze CO absorption for these targets.

\subsection{Absorption modeling} \label{sec:analysis:vpfit}

Absorption due to a CO isotope arises from different rotational energy levels.
We assume that the levels are thermalized and follow a Boltzmann distribution.
The absorption is then described by four parameters: a central velocity $v_{cen}$, a broadening parameter $b$, a total column density across all levels $N_{tot}$, and an excitation temperature $T_{ex}$.
We assume uniform priors on these parameters in either linear or logarithmic space:
\begin{align}
    v_{cen}/\text{km s}^{-1} &\sim \text{Uniform}\left(v_{min},\, v_{max}\right)\\
    \lten \left(b/\text{km s}^{-1}\right) &\sim \text{Uniform}\left(\lten 0.05,\, \lten 10\right)\\
    \lten \left( N_{tot}/\text{cm}^{-2} \right) &\sim \text{Uniform}(11, 18)\\
    T_{ex}/\text{K} &\sim \text{Uniform}(2.5, 20).
\end{align}
The velocity limits $v_{min}$ and $v_{max}$ extend $\pm$ 15 \kms~around the central velocity of the strongest feature seen in other molecular and dense neutral gas tracers in the spectrum.
The tracers we use are \ion{S}{1}, \ion{Cl}{1}, and \ion{C}{1}.

Before fitting, we determine whether \isoCO{12} absorption is detected by calculating the Bayes factor between a continuum-only model and a continuum and absorption model.
Specifically, the Bayes factor in this case is the marginal likelihood of the more complicated model divided by the marginal likelihood of the less complicated model.
The marginal likelihood of the continuum-only model is available directly, since the probabilistic part of the continuum model is a Gaussian process.
To estimate the marginal likelihood $p({\bf y})$ of the continuum and absorption model, we draw parameter sets $\boldsymbol{\theta}_i$ from a restricted version of the prior, calculate the likelihood, and average:
\begin{equation}
    p({\bf y}) \approx \frac{1}{N} \sum_{i=1}^{N} p({ \bf y}\vert \boldsymbol{\theta}_i).
    \label{eqn:marginal-likelihood}
\end{equation}
The restriction to the prior is that the $\lten$ column density is limited to be between 11 and 14, a range wide enough that the upper bound is clearly distinguishable from no absorption but narrower than the original to make sampling easier.

\autoref{eqn:marginal-likelihood} is the naive Monte Carlo estimator of the marginal likelihood and is known to be more of a lower bound when the likelihood is highly concentrated relative to the prior.
This limitation is fine for our purposes: we want to be able to distinguish between detections, which will have a concentrated likelihood, and non-detections, which will have a fairly diffuse likelihood (very roughly, a 4D hyperrectangle with an upper bound along the column density direction).
As a result, marginal likelihoods will be lower bounds for detections but reasonably accurate for non-detections.

There are differences of opinion on the exact Bayes factor at which one can claim a detection \citep{Kass:1995vb}, but a general consensus that a Bayes factor less than 10 is, at best, weak evidence while one greater than 100 is strong-to-decisive evidence for a detection.
In our case, there are 42 targets with Bayes factors less than eight, eight targets with a Bayes factor greater than 235, and none with values between eight and 235. 
We take the eight targets with values greater than 235 to be detections and the rest to be non-detections. 

To check if our procedure for detecting CO tends to produce false positives, we repeat all the same steps---stellar model fitting, Gaussian process parameter optimization, and Bayes factor estimation---for 20 ULLYSES sources with \molH\ column densities below $10^{19}$ \cmmt. 
Based on the Milky Way literature, lines of sight with \molH\ column densities this low should not contain detectable amounts of CO \citep{Sonnentrucker:2007ux}.
In this low \NmolH\ sample, the maximum Bayes factor in favor CO presence is 2.9.
All 20 low-\NmolH\ targets are therefore CO non-detections, raising our confidence that we are not claiming spurious detections of CO absorption.

\begin{figure*}
\centering
\includegraphics{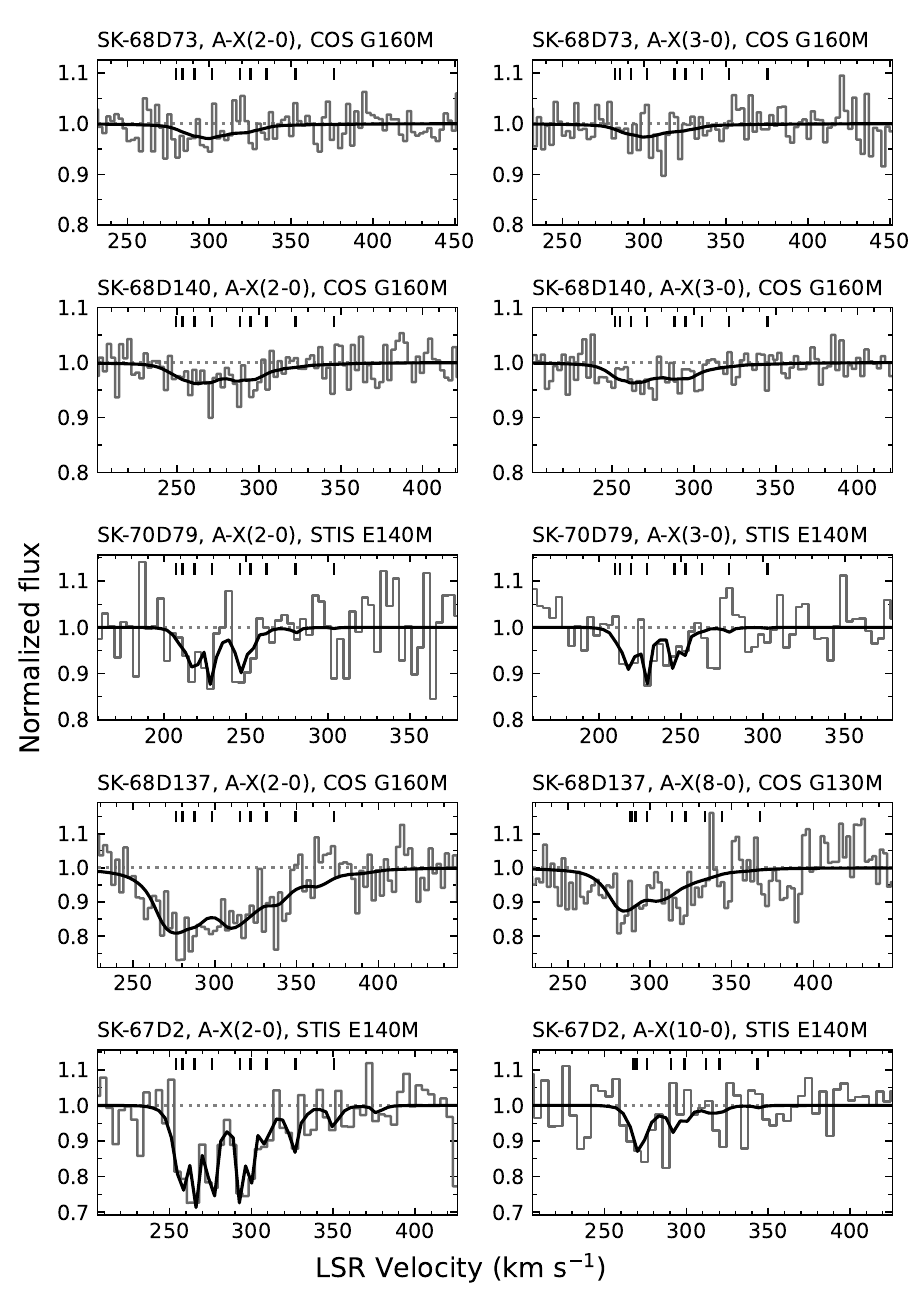}
\caption{Continuum-normalized spectra and \isoCO{12} absorption fits for targets in the Large Magellanic Cloud. The figure shows two CO bands, the strong A-X (2-0) band (left column) and a weaker band that varies from target to target (right column). The velocity scale is relative the rest wavelength of the chosen band's transition from the $J=0$ rotational state. Tick marks above the spectrum indicate the velocity centroids of transitions from $J$ less than 4. Note that the fits are done to all of the available CO bands in the spectra, not just the ones shown here.}
\label{fig:abs-fit-lmc}
\end{figure*}

\begin{figure*}
\centering
\includegraphics{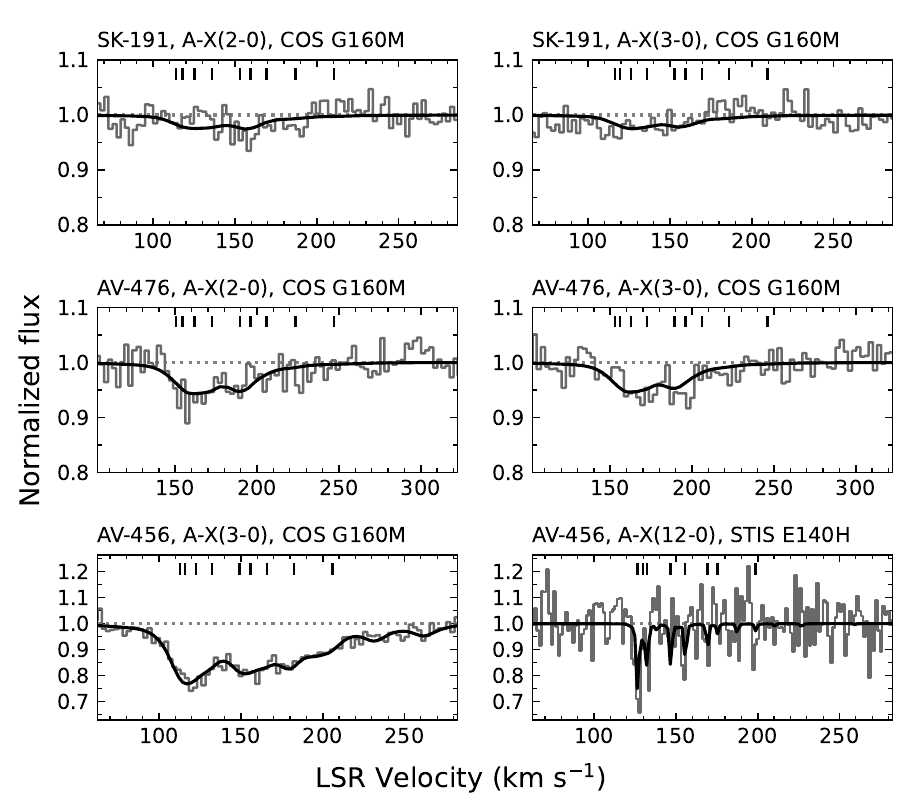}
\caption{Continuum-normalized spectra and \isoCO{12} absorption fits for targets in the Small Magellanic Cloud. The figure shows two CO bands, the strong A-X (2-0) band (left column) and a weaker band that varies from target to target (right column). The velocity scale is relative the rest wavelength of the chosen band's transition from the $J=0$ rotational state. Tick marks above the spectrum indicate the velocity centroids of transitions from $J$ less than 4. Note that the fits are done to all of the available CO bands in the spectra, not just the ones shown here.}
\label{fig:abs-fit-smc}
\end{figure*}

\begin{deluxetable*}{llccccccccc}
\tablecaption{Absorption parameters for LMC and SMC detections \label{tab:detections}}
\tablehead{
ULLYSES ID & Species & \colhead{$\lten{\NCO}$ low} & \colhead{$\lten{\NCO}$} & \colhead{$\lten{\NCO}$ high} & \colhead{$v_{cen}$} & \colhead{$\sigma$ $v_{cen}$}
 & \colhead{$\lten b$} & \colhead{$\sigma$ $\lten b$} & \colhead{$T_{ex}$} & \colhead{$\sigma$ $T_{ex}$}\\
 & & [\cmmt] & [\cmmt] & [\cmmt] & (\kms) & (\kms) & [\kms] & [\kms] & (K) & (K) 
}
\startdata
SK-68D73	& \isoCO{12} & 12.86 & 13.06 & 13.29 & 301.78 & 10.75 & -0.30	& 0.42	& 8.57	& 4.00 \\
SK-68D140	& \isoCO{12} & 13.14 & 13.46 & 13.53 & 271.33 & 4.54 & -0.31	& 0.47	& 12.15	& 4.24 \\
SK-70D79	& \isoCO{12} & 13.39 & 13.58 & 13.67 & 229.21 & 0.55 & -0.12	& 0.33	& 7.81	& 2.29 \\
SK-68D137	& \isoCO{12} & 15.08 & 15.26 & 15.39 & 289.34 & 1.18 & -0.33	& 0.04	& 8.84	& 1.39 \\
SK-67D2		& \isoCO{12} & 15.61 & 15.93 & 16.54 & 275.88 & 0.46 & -0.64	& 0.19	& 5.86	& 1.09 \\
SK-67D2		& \isoCO{13} & 13.46 & 14.10 & 14.21 & 275.79 & 0.53 & -0.34	& 0.30	& 4.23	& 1.37 \\
SK-191		& \isoCO{12} & 13.17 & 13.39 & 13.58 & 135.96 & 2.22 & -0.78	& 0.40	& 11.21	& 2.96 \\
AV-456		& \isoCO{12} & 15.95 & 16.04 & 16.13 & 132.39 & 0.06 & -0.46	& 0.05	& 17.12	& 1.43 \\
AV-456		& \isoCO{13} & 14.58 & 14.82 & 14.95 & 132.03 & 0.35 & -0.67	& 0.10	& 15.38	& 3.23 \\
AV-476		& \isoCO{12} & 13.41 & 13.64 & 13.76 & 172.46 & 1.73 & -0.37	& 0.32	& 9.64	& 2.40 \\
\enddata
\tablecomments{Absorption parameters and uncertainties for \isoCO{12} and, when it is detectable, \isoCO{13}. Total column densities are reported as a mean and a 68\% highest posterior probability density interval. The remaining parameters are reported as means and standard deviations.}
\end{deluxetable*}

Upper limits on the $\lten$ column densities for the 42 high-\NmolH\ sample non-detections are reported in an Appendix.
To measure column densities for the detections, we use a Markov chain Monte Carlo (MCMC) implementation in \texttt{numpyro} to sample the posterior probability density function over absorption parameters. 
The continuum can be analytically marginalized over because the probabilistic part of the continuum model is a Gaussian process.
We first run MCMC assuming only \isoCO{12} is present, visually inspect the fits for the presence of un-modeled \isoCO{13} absorption, and re-run MCMC with an additional (and independent) set of \isoCO{13} parameters if necessary.

Figures \ref{fig:abs-fit-lmc} and \ref{fig:abs-fit-smc} show normalized spectra and absorption fits for the eight detections. 
The figures show two CO bands per target: the strong A-X(2-0) band and a weaker band that varies from target to target. 
In every case, other CO lines that are not shown are also used in the fit.
We report medians and central 68\% intervals of the parameter posterior distributions for the detections in \autoref{tab:detections}.
Our column density uncertainties can be large.
This is particularly true at higher column densities---the 68\% interval for \sknum{67}{2}{, for example, spans almost an order of magnitude.
These large column density uncertainties are an unfortunate consequence of deep absorption features that are much narrower than the instrumental LSF.
To give a sense of how important unresolved saturation is in the case of} \sknum{67}{2}, the maximum oscillator strength for the $A-X(2-0)$ band shown in the left panel of the fit figure is 100 times greater than the maximum oscillator strength for the $A-X(10-0)$ band shown in the right panel but the $A-X(2-0)$ feature is only about 30\% deeper than the $A-X(10-0)$ feature.
In this saturation regime, there is considerable degeneracy between line width and column density. 
This degeneracy is responsible for the large uncertainties.

The fitting procedure needs to be validated since it is a new implementation that has not been used before.
To test the consistency of its outputs with prior work, we measure \isoCO{12} column densities for six Milky Way sightlines from \citet{Sheffer:2008vc}: HD 14434, HD 94454, HD 116852, HD 137595, HD 144965, and HD 308813.
These lines of sight were chosen to cover a broad range in previously reported total column density ($10^{13.30}$ to $10^{15.28}$ \cmmt) and to include a case with multiple components (HD 14434).
We note that we do not expect our total column densities to exactly match the ones published in \citet{Sheffer:2008vc} because we are using a different reduction of the spectra, a different set of CO oscillator strengths, and different LSFs (tables from the HST website in our case, a Gaussian with width a fit parameter in their case).

To summarize the test results, we calculate the root mean square error (based on the difference between our point estimate and their point estimate), mean of the standardized residuals (our estimate minus their estimate over standard deviation) and the chi-squared statistic.
The standard deviation here is the sum in quadrature of the 20\% uncertainty stated as a conservative estimate in \citet{Sheffer:2008vc} and the standard deviation we calculate using our fitting procedure.
The measurements are not truly independent (they use the same observations, though with different reductions), so these summary values provide a general sense of agreement or disagreement but should not be interpreted too rigorously.

The new (literature) logarithmic column densities for these lines of sight, listed in the same order as the star names, are: 14.43 (14.36), 14.43 (14.3), 13.25 (13.3), 13.92 (13.89), 15.19 (15.28), and 13.79 (13.84). 
The root mean square error is 0.08 (21\%), the mean of standardized residuals is 0.07 and the chi-squared statistic is 4.3 corresponding to a chi-squared tail probability of 63\% (meaning a 63\% probability of observing a statistic at least as large as 4.3). 
We see an acceptable level of agreement and no obvious differences in performance across the column density range.

\section{Results} \label{sec:results}

\begin{figure*}
    \centering
    \includegraphics[width=\linewidth]{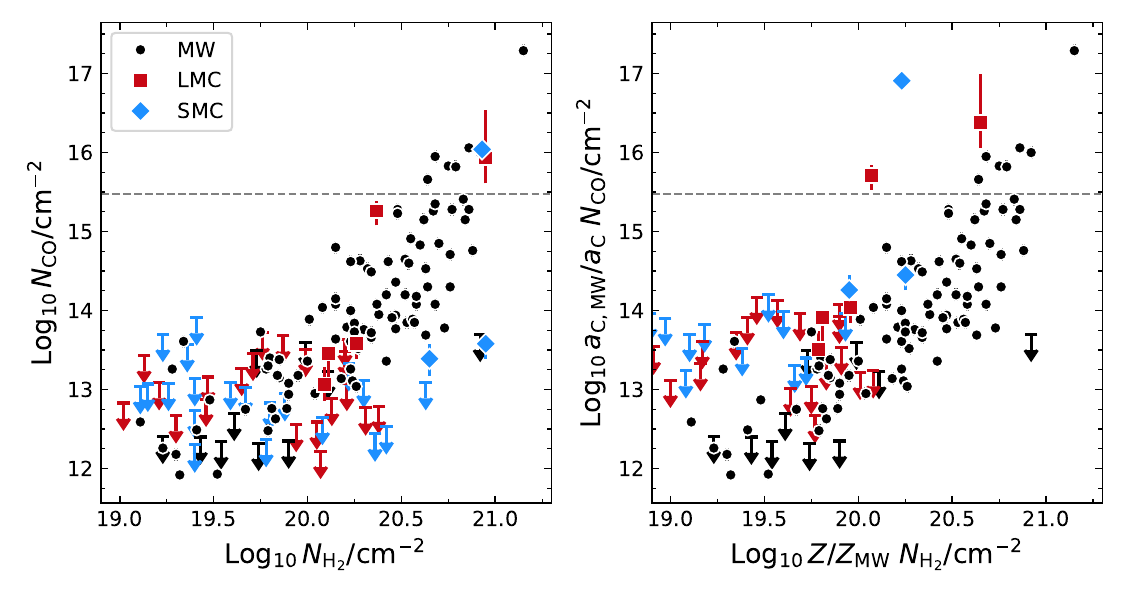}
    \caption{Measured (left) and abundance-rescaled (right) \isoCO{12}\ and \molH\ column densities in the Milky Way (black), Large Magellanic Cloud (red; LMC), and Small Magellanic Cloud (blue; SMC). Milky Way values are taken from the literature (see beginning of \S \ref{sec:results}), LMC and SMC values are from this work. Upper limits on non-detections are denoted by downward facing arrows of the appropriate color. Detections are denoted by circles, squares, and diamonds, respectively. The horizontal dashed gray line is an estimated column density threshold for CO to be detected in emission in current surveys of the LMC and SMC. The left panel shows the actual measurements. At fixed \molH\ column density, CO column densities tend to be lower in the LMC and SMC relative to the Milky Way. The right panel ``reverses'' a hypothetical effect of metallicity on the \NCO(\NmolH) relation. \NCO\ is assumed to shift ``vertically'' due to the decrease in carbon abundance $a_{\rm C}$ and to shift ``horizontally'' due to the smaller dust optical depth $A_V$ per hydrogen nucleus. Reversing both shifts results in an overcorrection: the abundance-rescaled CO column densities for the LMC and SMC are clearly higher than the Milky Way column densities.}
    \label{fig:H2-CO-data}
\end{figure*}

The left panel of \autoref{fig:H2-CO-data} shows UV absorption measurements of \NCO~as a function of \NmolH~for the Milky Way, LMC, and SMC.
The Milky Way measurements are taken from the literature \citep{Rachford:2002wo,Crenny:2004ul,Burgh:2007vc,Sonnentrucker:2007ux,Sheffer:2008vc,Burgh:2010wd} while the Magellanic measurements come from this work.
The difference in \NCO\ as a function of \NmolH\ between the Milky Way and MCs can be thought of as the result of two shifts: a vertical one due to a change in carbon abundance and a horizontal one due to a change in processes such as shielding by dust.

\citet{Wolfire:2010-CO-dark-gas-layer} predict that the difference in dust optical depth $\Delta A_V$ between a \molH\ cloud boundary $R_{\molH}$ and a CO cloud boundary $R_{\rm CO}$ depends only weakly on the radiation field and gas metallicity.
That $\Delta A_V$ would correspond to a hydrogen column of $GDR \times \Delta A_V$, where $GDR$ is the gas-to-dust ratio.
The $GDR$ declines at least linearly with decreasing metallicity.
The separation in hydrogen column density between $R_{\molH}$ and $R_{\rm CO}$ should therefore grow by at least a factor of $Z_\odot/Z$ as the metallicity declines.
This would imply a horizontal shift in $\NCO(\NmolH)$ of about $Z_\odot/Z$ to the right.

The right panel of \autoref{fig:H2-CO-data} shows the result of ``reversing'' a horizontal shift by $Z_\odot/Z$ and a vertical shift by the carbon abundance.
This combination of shifts is too large.
The LMC and SMC points at re-scaled $\NmolH\approx 10^{20.25}$ \cmmt\ are substantially above the actual Milky Way measurements.
The SMC value is, in fact, close to the total carbon column density for its \NmolH.

The lack of a horizontal shift suggests that the relative sizes of the CO-rich and \molH-rich parts of molecular clouds do not change dramatically between solar and $1/5$ solar metallicity.
This lack of ``CO shrinkage'' in turn implies a relatively shallow dependence of $X_{CO}$ on metallicity. 
Below, we quantify how the \NCO(\NmolH) relation changes from the Milky Way to the Magellanic Clouds.
We then use then \NCO(\NmolH) relation to estimate the metallicity dependence of the fraction of \molH\ mass with detectable CO emission.

\subsection{Metallicity dependence of the $\NCO(\NmolH)$ relation}
\label{sec:results-NCO-NH2}

The distribution of \NCO\ as a function \NmolH\ for the Milky Way sample has two main characteristics: a change in the slope of the $\lten \NCO$ as a function of $\lten \NmolH$\ relation at about 10$^{20.4}$ \cmmt\ \citep{Sheffer:2008vc} and scatter in \NCO\ at fixed \NmolH\ that is larger than the measurement uncertainties. 
Both characteristics are the result of simple physical and chemical processes and so should also apply in the LMC and SMC.
However, in these galaxies there are too few measurements and too many upper limits below the knee in the relation to independently model each galaxy's slope change and scatter.
We instead model all three galaxies together, with some parameters being shared while others are allowed to vary from galaxy to galaxy.

The model is based on the assumption that the fraction of carbon in the form of CO at a distance $y$ along a line of sight, $f_{CO}(y)$, is a logistic function of the cumulative \molH\ column density along the line of sight \NmolH$(y)$:
\begin{equation}
    f_{CO}(\NmolH(y)) = \frac{1}{1 + \exp\left(- \frac{\NmolH(y) - x_0}{s} \right)}.
\end{equation}
$x_0$ is the \NmolH\ value where $f_{CO}$ reaches $1/2$ and $s$ is a ``slope'' that controls the steepness of the rise in $f_{CO}$ with $\NmolH$.
$f_{CO}$ does not have to specifically be the logistic function.
Other sigmoid (``s-shaped'') functions that decline slowly towards small $\NmolH$ yield similar shapes of $\NCO(\NmolH)$. 
However, steep functions such as power laws in $\NmolH$\ or sigmoid functions of $\lten\NmolH$ cannot reproduce the shallow slope of $\NCO(\NmolH)$ below the ``knee'' in the relation.

The column density corresponding to this $f_{CO}$ can be found by integrating $f_{CO}$ over $\NmolH$:
\begin{equation}
    \NCO(X) = 2 a_C \int_{0}^{X} f_{CO}(X') \, {\rm d} X',
\end{equation}
where $2 a_C$ is the carbon abundance per hydrogen molecule.
For our choice of $f_{CO}$, this integral evaluates to:
\begin{equation}
    \NCO(X) = 2 a_C \left[ \ln \left( e^{(X-x_0)/s} + 1 \right) - \ln \left( e^{-x_0/s}  \right) \right].
\end{equation}

We introduce sightline-to-sightline scatter by allowing a galaxy to have a range of $x_0$ values, where $x_0$ is log-normal about a per-galaxy mean $x_\mu$ with standard deviation $\sigma_x$.
This scatter term reflects a combination of uncertainty in \NmolH\ measurements and real variation of conditions in molecular clouds.
We set $\sigma_x$ and the relation slope $s$ to be the same across sightlines and galaxies.
It is not possible to allow sightline-to-sightline variation in both $x_0$ and $s$, as this would make them perfectly degenerate, and the LMC and SMC do not have enough informative data to independently constrain $\sigma_x$ and $s$.
If we do allow the LMC and SMC $\sigma_x$ and $s$ to vary, the inferred values have broad uncertainties and are consistent with the value for the Milky Way.

We approximate the likelihood as being normal in $\lten \NCO$.
We define the likelihood of upper limits using their stated coverage probability.
For example, a model $\NCO$\ that is less than a $2\sigma$ upper limit will have likelihood 0.95.

We implement this model and sample from the resulting posterior probability distribution using an MCMC sampler in \texttt{numpyro}.
To improve mixing of MCMC, we sample not in $x_\mu$ and $x_0$ but in $b_\mu=x_\mu/s$ and $b_0=x_0/s$.
This does not affect $\sigma_x$, since division in linear space is subtraction in logarithmic space.
The full set of priors is:
\begin{equation}
    \begin{aligned}
    \lten s &\sim \text{Uniform}(18, 22) \\
    \sigma_x &\sim \text{Gamma}(2, 2)\\
    \lten b_\mu &\sim \text{Uniform(-5, 5)} \\ 
    \lten b_{0,i} &\sim \text{Normal}(\lten b_\mu, \sigma_x^2),
    \end{aligned}
\end{equation}
where the Gamma distribution is defined in terms of shape and rate parameters.
The Gamma(2, 2) prior keeps $\sigma_x$ away from zero and has mean 1 and standard deviation $1/\sqrt{2} \approx 0.7$.
Avoiding $\sigma_x=0$ is necessary because having all $b_{0,i}$ values equal to any $b_\mu$ gives a finite (though possibly small) likelihood, but an infinite prior probability.
The log-uniform priors on $s$ and $b_\mu$ are broad enough to span the full range of plausible values. 
$\NCO(\NmolH)$ relations with $s$ or $b_0=b_\mu$ at the boundary of this range are clearly visually inconsistent with the measurements.

\begin{deluxetable}{lcc}
\tablecaption{$\NCO(\NmolH)$ model parameters \label{tab:NCO-NH2-pars}}
\tablehead{
\colhead{Parameter} & \colhead{Mean} & \colhead{Standard deviation}
}
\startdata
$\lten s/\text{\cmmt}$ & 20.080 & 0.040 \\
$\lten b_\mu$ MW & 0.917 & 0.011 \\
$\lten b_\mu$ LMC & 0.956 & 0.028 \\
$\lten b_\mu$ SMC & 1.020 & 0.037 \\
$\sigma_x$ & 0.079 & 0.006 \\
\tableline
$\lten x_\mu/\text{\cmmt}$ MW & 20.997 & 0.033 \\
$\lten x_\mu/\text{\cmmt}$ LMC & 21.036 & 0.045 \\
$\lten x_\mu/\text{\cmmt}$ SMC & 21.100 & 0.043 \\
\enddata
\tablecomments{Means and standard deviations of the posterior probability distribution over the parameters of the $\NCO(\NmolH)$ model described in \S\ref{sec:results-NCO-NH2}. $\text{Log}_{10} s$, $\lten b_\mu$, and $\sigma_x$ are the parameters used in the fit. $\text{Log}_{10} x_\mu$ is the sum of $\lten s$ and $\lten b_\mu$.}
\end{deluxetable}

\begin{figure*}
    \centering
    \includegraphics[width=\linewidth]{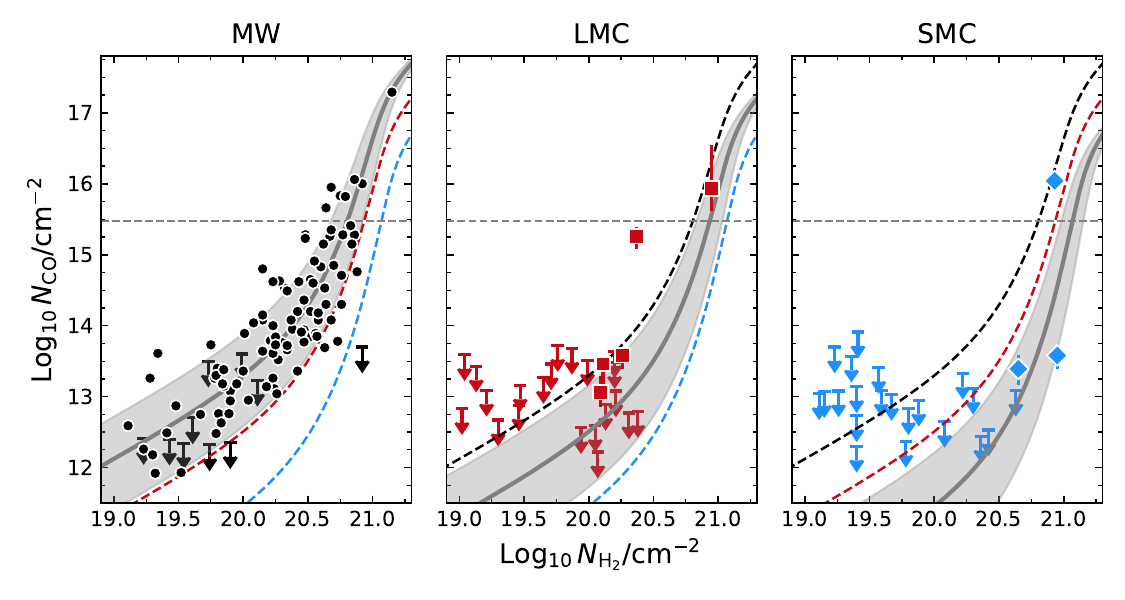}
    \caption{CO and \molH\ column density measurements and \NCO(\NmolH) model fits for the Milky Way, Large Magellanic Cloud (LMC), and Small Magellanic Cloud (SMC). Milky Way measurements are taken from the literature (see beginning of \S \ref{sec:results}), LMC and SMC values are from this work. Points and downward arrows are measurements and upper limits on non-detections. The horizontal dashed gray line is an estimated column density threshold for CO to be detected in emission in current surveys of the LMC and SMC. The solid gray line is the median \NCO(\NmolH) fit for the galaxy whose data is shown in a panel. The dashed black (Milky Way), red (LMC), and blue (SMC) lines are the medians for the other galaxies. The model includes scatter around the median relation. This scatter is represented by the gray shaded regions. The figure does not show uncertainty in fit parameters. As the metallicity declines, the \NCO(\NmolH) relation shifts down and to the right.}
    \label{fig:fit-NCO-NH2}
\end{figure*}

Fit parameters and uncertainties are listed in \autoref{tab:NCO-NH2-pars}.
\autoref{fig:fit-NCO-NH2} shows the model evaluated at the mean parameters from the table.
Each panel includes the measurements for a galaxy, the estimated median \NCO(\NmolH) relation for that galaxy (solid gray line), the median relations for the other two galaxies (dashed black, red, and blue lines), and the $\pm 1\sigma_x$\ width of the model due to intrinsic scatter in the $x_0$ parameter (gray region). 
Note that this gray region does not represent model uncertainty.

The median SMC relation is lower than that of the LMC, which is lower than that of the Milky Way. 
The scatter in \NCO\ at fixed \NmolH\ is large, leading to overlap between the galaxies' distributions of \NCO\ at a constant value of \NmolH. 
While the scatter model is not explicitly physical, it likely reflects the combined effect of density and radiation field intensity variations within a galaxy. 
The overlap of \NCO(\NmolH) distributions between the Milky Way and LMC shows that the amplitude of these fluctuations can affect CO abundances by about as much as a 50\% reduction in metallicity.

Our \NCO(\NmolH) function can be shifted vertically by changing the carbon abundance and (mostly) horizontally by changing $x_\mu$. 
If the CO abundance at at fixed \NmolH\ depends non-linearly on metallicity, then $x_\mu$ should change from the Milky Way to the lower-metallicity Magellanic Clouds.
$\lten x_\mu /$\cmmt\ in the SMC is about 2.7$\sigma$\footnote{Note that the uncertainties on $\lten x_\mu/$\cmmt\ for the three galaxies are correlated, so the standard deviation of the difference between two of them is smaller than the standard deviation on an individual value.} lower than it is in the Milky Way, confirming with intermediate statistical significance that CO chemistry depends non-linearly on metallicity.
However, the difference in $\lten x_\mu /$\cmmt\ is smaller than the difference in metallicity. 
The difference of the point estimates for the Milky Way and SMC is $-0.1$, a conservative assumption of 5$\sigma$ would give $-0.2$, and the difference in $\lten$\ metallicity is $-0.7$. 
The \NCO(\NmolH) relation does, as expected, shift along the \NmolH\ axis, but by less than the metallicity.
This quantitative analysis confirms the qualitative argument from the beginning of this section.

\subsection{Metallicity dependence of the CO-bright \molH\ fraction}
\label{sec:results-CO-bright-frac}

The conversion between CO column density and intensity \letCO{I}\ depends on conditions in the gas and requires radiative transfer modeling. 
Since we do not have a model of gas conditions, we cannot directly estimate \letCO{X}\ from the $\NCO(\NmolH)$ relations.
In future work, we will use the \NCO\ measurements to evaluate the accuracy of models and simulations that predict both $\NCO(\NmolH)$ and \letCO{X}.
This comparison will provide a model-dependent estimate of \letCO{X}. 

A quantity that we can estimate from the current observations and a few additional assumptions is the fraction of \molH\ associated with detectable CO emission.
Explicitly, this fraction is the mass of hydrogen with CO intensity \letCO{I} above a detection threshold divided by the total mass of hydrogen.
This quantity will be denoted as $f_b$.
To estimate $f_b$ from $\NCO(\NmolH)$, we need to define a threshold column density, $N_{thresh}$, above which CO emission will typically be detected.
In general, this threshold will depend on the CO transition, the size of CO clouds, and the sensitivity and spatial resolution of the observations.
We will pick $N_{thresh}$ using existing surveys of CO emission in the LMC and SMC.

The two highest-\NCO\ sightlines in the LMC, \sknum{68}{137} and \sknum{67}{2} are in the footprint of the MAGMA survey of the $J=1-0$ transition \citep{Wong:2017-MAGMA-DR3}.
These two LMC sightlines and the highest \NCO\ SMC sightline, AV 456, are in the footprint of the APEX $J=3-2$ transition surveys of the LMC and SMC (\citealt{Grishunin:2023-APEX-LMC-initial}; Chen \& Weiß priv. comm.).
These surveys have beam full widths at half maximum of about 11 pc and 5 pc, respectively.
Neither line is detected toward \sknum{68}{137}; $J=1-0$ emission is detected at the position of \sknum{67}{2} and $J=3-2$ emission is detected one 1.5 parsec pixel away; $J=3-2$ emission is detected at the position of AV 456.
The detection threshold should therefore be between the \NCO\ of the emission non-detection \sknum{68}{137} and the \NCO\ of the emission detections \sknum{67}{2} and AV 456.
We will use $N_{thresh} = 3 \times 10^{15}$ \cmmt.

\begin{figure}
    \centering
    \includegraphics[width=\linewidth]{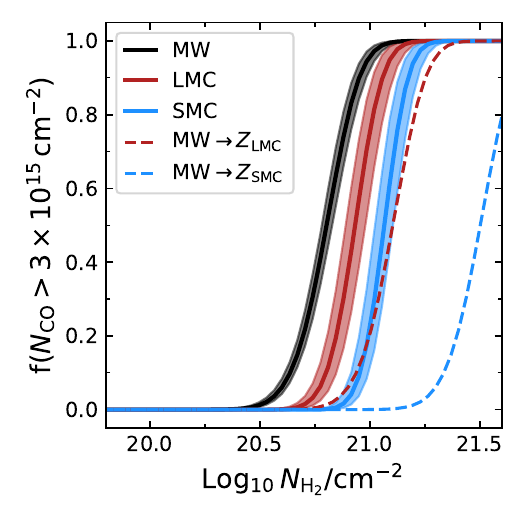}
    \caption{Expected fraction of sightlines at a given \molH\ column density that will have a CO column density greater than $3\times10^{15}$ \cmmt. This threshold is an estimate for the column density above which CO emission is detected in current surveys of the Large and Small Magellanic Clouds (LMC and SMC). Solid black (Milky Way), red (LMC), and blue (SMC) lines are median fractions derived from the \NCO(\NmolH) models shown in \autoref{fig:fit-NCO-NH2} and described in \S\ref{sec:results-NCO-NH2}. Shaded regions represent $1\sigma$\ uncertainties on the parameters of those models. Dashed lines are the Milky Way fraction shifted right by the inverses of the metallicities of the LMC and SMC: $f_{MW}(\NmolH\times Z_{MW}/Z_{MC})$. Taking a reference value of the fraction, for example 50\%, as the CO-bright cloud edge, the dashed lines correspond to a scenario where the cloud edge changes by a factor of the metallicity. We find a smaller shift, implying that the CO-bright parts of molecular clouds shrink relatively slowly as the metallicity declines.}
    \label{fig:f-bright-at-h2}
\end{figure}

The $\NCO(\NmolH)$ model described in \S\ref{sec:results-NCO-NH2} assigns a distribution of \NCO\ values to each \NmolH\ value. 
The integrated probability above $N_{thresh}$ at some \NmolH\ can be thought of as the bright fraction at a particular \molH\ column density, $f_b(\NmolH)$.
\autoref{fig:f-bright-at-h2} shows $f_b(\NmolH)$ for the Milky Way, LMC, and SMC.
The uncertainties are estimated by calculating $f_b(\NmolH)$ for each sample from the posterior probability distribution over model parameters and taking the 16th and 84th percentiles of $f_b$ at each \NmolH. 
The CO-bright zone moves deeper into the cloud as the metallicity declines, but the amplitude of this shift is less than the change in metallicity.

By combining $f_b(\NmolH)$ with a realistic \molH\ column density distribution, we can estimate the CO-bright mass fraction for a hypothetical molecular cloud.
Given the distribution of \molH\ mass as a function of \NmolH, ${\dd M_{\molH}}/{\dd \NmolH}$, the CO-bright mass fraction is the integral over \NmolH\ of $f_b(\NmolH) \times {\dd M_{\molH}}/{\dd \NmolH}$.
To have a realistic molecular hydrogen distribution, we use a chemically post-processed snapshot of one of the TIGRESS simulations\footnote{Obtained from the TIGRESS public data release: \href{https://princetonuniversity.github.io/astro-tigress/intro.html}{princetonuniversity.github.io/astro-tigress/intro.html}.} \citep{Kim:2017-TIGRESS-general-ref,Gong:2018-TIGRESS-XCO-fiducial,Gong:2020-TIGRESS-XCO-variation}.
This is a simulation of a kiloparsec by kiloparsec region in a Milky Way metallicity disk galaxy.

\begin{figure*}
    \centering
    \includegraphics[width=\linewidth]{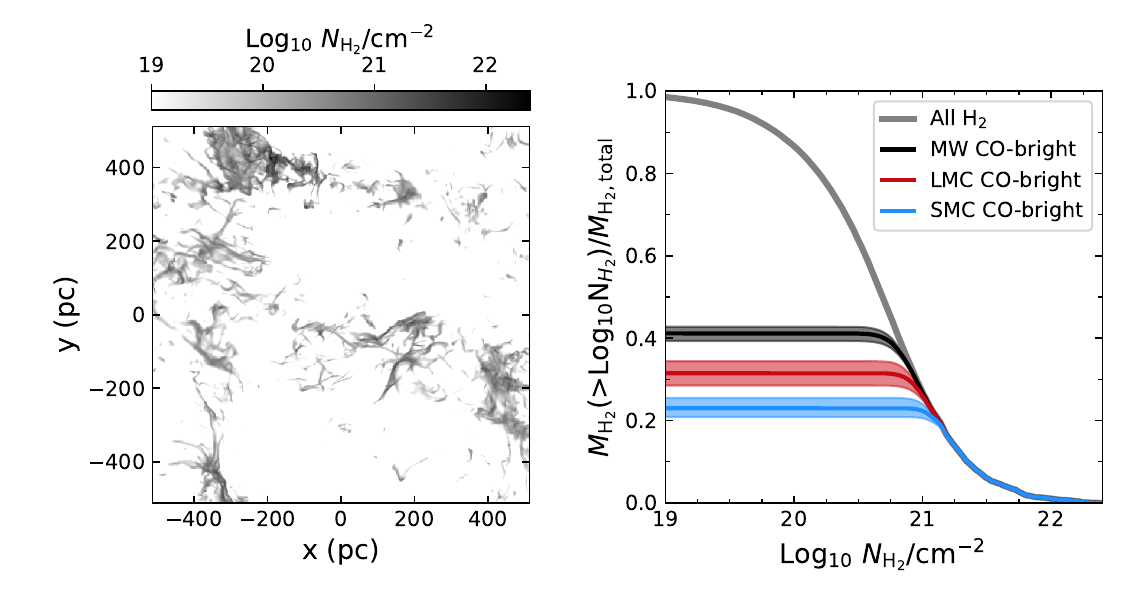}
    \caption{\molH\ column density distribution (left) and complementary cumulative mass distribution function (right) computed from an interstellar medium simulation \citep{Gong:2020-TIGRESS-XCO-variation}. The grey curve in the right panel shows the fraction of \molH\ mass found along lines of sight with \NmolH\ greater than a given threshold value in the left panel. The black, red, and blue curves in the right panel show this \molH\ mass fraction weighted by the probability of detecting CO emission as a function of \NmolH\ in the Milky Way, Large Magellanic Cloud, and Small Magellanic Cloud, respectively. The detection probability curves are shown in \autoref{fig:f-bright-at-h2} and were derived assuming a \NCO\ detection threshold of $3\times 10^{15}$ \cmmt. The total CO-bright mass fraction calculated from this particular \NmolH\ distribution declines by about a factor of two from the Milky Way to the Small Magellanic Cloud, while the metallicity declines by a factor of about five.}
    \label{fig:f-bright-over-distr}
\end{figure*}

We project the \molH\ density along the box height direction to get a column density map, which is shown in the left panel of \autoref{fig:f-bright-over-distr}.
The right panel of the figure shows the cumulative mass fraction of total or CO-bright \molH\ mass as a function of \molH\ column density.
For the \molH\ distribution in this simulated region, the median (with respect to \NCO(\NmolH) model uncertainties) CO-bright mass fraction is 0.41, 0.32, and 0.23 for the Milky Way, LMC, and SMC.
The ratios of the CO-bright mass fractions to that of the Milky Way, $f_{b, MC}/f_{b, MW}$, are 0.70 to 0.83 for the LMC and 0.51 to 0.61 for the SMC.
The metallicities of the LMC and SMC are 0.5 and 0.2 times that of the Milky Way, respectively, meaning that the CO-bright mass fraction for this \molH\ distribution declines more slowly than the metallicity.

The CO bright fraction depends on a cloud's \NmolH\ distribution. 
The clouds in the top right quadrant of the left panel of \autoref{fig:f-bright-over-distr} have a lower peak \NmolH\ than the region as a whole: $10^{22.36}$ \cmmt\ across the region, $10^{21.12}$ \cmmt\ for this quadrant. 
The CO bright fraction for this quadrant based on the Milky Way $f_b(\NmolH)$ is 0.14, and the ratio of bright fractions relative to the Milky Way is 0.34 to 0.60 for the LMC and 0.07 to 0.20 for the SMC. 
For a particularly diffuse molecular cloud, the CO bright fraction is still consistent with an approximately linear dependence on metallicity.

A linear or sub-linear relation between metallicity and $f_b$ suggests a similar dependence of \letCO{X} on metallicity. 
There is a small difference in column density between the point where CO emission becomes detectable and the point where it becomes optically thick. 
\letCO{X} at the resolution of a cloud is then roughly proportional to the cloud-scale CO bright fraction.
The analysis in this section therefore implies that \letCO{X} depends linearly or sub-linearly on metallicity.
We will make a more direct estimate of $\letCO{X}(Z)$ in future work.

\section{Discussion} \label{sec:discussion}

\subsection{Limitations and consequences} \label{sec:discussion:limitations}
One limitation of this work is the low spectral resolution compared to the width of CO features.
Because of the complex shape of the stellar continuum, it is difficult to distinguish weak CO absorption from continuum variation at COS or even STIS E140M resolution. 
This problem is the cause of our relatively high upper limits on non-detections. 

The main consequence of the weak upper limits is that we do not know how much lower \NCO\ in the low-\NmolH\ regime is in the LMC and SMC compared to the Milky Way.
Below $\approx 10^{20}$ \cmmt, the LMC and SMC \NCO(\NmolH) relations from \S \ref{sec:results-NCO-NH2} are, effectively, extrapolating from higher \NmolH\ at the same metallicity and from Milky Way metallicity at the same \NmolH.
This is not important for understanding \letCO{X}, since CO emission at the corresponding \NCO\ is too faint to detect even in the Milky Way.
It would, however, have been a useful piece of information for understanding molecular chemistry in the diffuse molecular regime \citep{Zsargo:2003-nonthermal-CHplus,Visser:2009-CO-isotopologue-model,Gerin:2021-COplus-CO}.

A potentially more important limitation is the sample definition.
The sample is small in general and particularly small at $\NmolH > 10^{20.4}$ \cmmt, the point above which the the Milky Way \NCO(\NmolH) relation steepens.
There are only six sightlines in the LMC and SMC combined at these \molH\ column densities.
In the Milky Way, \NCO\ at fixed high \NmolH\ varies by over an order of magnitude.
With our sample size, we cannot tell if this distribution changes from the Milky Way to the LMC or SMC.

The targets, massive UV bright stars, are also a biased population in a variety of ways.
We are studying the conditional distribution of \NCO\ given \NmolH.
This means that selection effects that alter the distribution of \NmolH\ without affecting the conditional distribution of \NCO\ will not affect the accuracy of the results. 
The association of young stars with molecular clouds and the bias against observing highly obscured sources should both act primarily on the distribution of \NmolH\ and are therefore not a concern. 

The contribution of the targets and any other clustered UV bright stars to the photodissociation rate of the foreground gas, however, could affect the conditional distribution of \NCO.
CO is less effective at self-shielding than \molH.
It is therefore expected to be more strongly suppressed than \molH\ as the strength of the radiation field increases.
This may be more of a problem in the LMC and SMC than in the Milky Way. 
Background sources in the Milky Way may be in a different part of the Galactic plane than the foreground molecular gas.
In the more face-on LMC and SMC, they probably are near the foreground gas.
We therefore provide the caveat to our results that the distribution of \NCO\ at fixed \NmolH\ could be higher for randomly chosen sightlines (or in an emission survey) than in our study.

Curiously, the sightline in the LMC with the highest \NCO\ is toward a probable hypervelocity runaway star, \sknum{67}{2} \citep{Lennon:2017-sk-67-2-hypervelocity}. 
This star may not actually be near the molecular gas in front of it.
Even if it is, the star's high speed means that it has only been there for a short time.

\subsection{Comparison with Magellanic CO absorption measurements from the literature} \label{sec:discussion:prev-measurements}
To our knowledge, detections of CO absorption in the ISM of the LMC and SMC have been claimed in four papers.
\citet{Koenigsberger:2001-implausible-CO-detection} analyze a STIS spectrum of the SMC star HD 5980 and state that they detect CO absorption at SMC velocities.
We do not see this absorption in the spectrum. 
The presence of CO along this sightline would be highly unexpected given that the SMC \molH\ column density toward HD 5980 is only $10^{15.66}$ \cmmt \citep{Tumlinson:2002tz}. 

\citet{Bluhm:2001aa-CO-Sk-69-246} and \citet{Andre:2004aa} both detect CO in a FUSE spectrum of \sknum{69}{246}, though with different column densities and uncertainties. 
The upper limit we derive from an analysis of a STIS E140M spectrum of this star is consistent with \citet{Bluhm:2001aa-CO-Sk-69-246} but below the measurement of \citet{Andre:2004aa}.
\citet{Andre:2004aa} also detect CO in two other FUSE spectra of LMC stars at levels that are inconsistent with our upper limits based on STIS E140M spectra.
We believe the explanation for the discrepancy is an issue noted in both works: the CO band they used is blended with \molH\ and \specnotation{Cl}{I} absorption and separating these features out is difficult at the resolution of FUSE.
The strongest CO bands analyzed in this work have line strengths $f\lambda$ that are about half that of the strong $C-X\,(0-0)$ band covered by FUSE, but have the benefit of not being blended with other ISM features.

\citet{Welty:2016-MC-thermal-pressures} note that CO absorption is present in the spectra of AV 456 and \sknum{68}{73}, but do not provide further analysis.
We measure significant CO absorption toward both targets as well.

\subsection{The \isoCO{12} to \isoCO{13} ratio}

The \isoratio{CO}{12}{13} abundance ratio is determined by a combination of nucleosynthetic and chemical processes.
Nucleosynthesis sets the \isoratio{C}{12}{13} ratio.
The \isoratio{CO}{12}{13}\ ratio can be greater than that of carbon because of isotope fractionation: isotopologues have different transition energies and so do not completely cross-shield each other from photodissociation \citep{Bally:1982-CO-isotope-photodestruction}.
More abundant isotopes will be more effective at self-shielding, leading to a ``rich get richer" situation.
When there is enough shielding from other sources, \isoratio{CO}{12}{13}\ can drop below the ratio for carbon due to isotope exchange reactions \citep{Watson:1976-CO-fractionation,Roueff:2015-fractionation-reactions}.
\isoCO{X} can react with $^{Y}$C$^+$ to form \isoCO{Y} and $^{X}$C$^+$.
This reaction is energetically favorable for \isoCO{12} and $^{13}$C$^+$, leading to an enhancement in the \isoCO{13} abundance when carbon ions are plentiful.

We measure \isoratio{CO}{12}{13}\ ratios toward two targets, \sknum{67}{2} (2 to 148; 68\% highest density interval) and AV 456 (10 to 25). 
\isoratio{C}{12}{13}\ is estimated to be 49-60 in the LMC \citep{Chin:1999-LMC-12C-13C,Wang:2009-LMC-12C-13C} and, to the best of our knowledge, has not been measured in the SMC.
The measurement for \sknum{67}{2} is too uncertain to be informative.
\citet{Szucs:2014-CO-fractionation} predict that a sightline with $N_\text{\isoCO{12}} \approx10^{15.8}$ \cmmt\ will have \isoratio{CO}{12}{13}$\approx 20$, which is consistent with the AV 456 measurement.
This prediction was made assuming an intrinsic carbon isotope ratio to 60.

The ratio for AV 456 is lower than most ultraviolet Milky Way measurements \citep{Sheffer:2007-12CO-13CO,Sonnentrucker:2007ux} but consistent with millimeter absorption ones \citep{Liszt:2017-CO-fractionation}.
If the AV 456 ratio is driven by chemical processes, the low value suggests that this cloud is in the intermediate-shielding regime: \isoCO{13}\ is not being efficiently photodissociated, but there are still enough $^{13}$C$^+$ ions for the exchange reaction to be important.
The optical depth to local sources of photodissociating radiation for this regime is expected to be $A_V\approx 2-3$ \citep{Warin:1996-CO-fractionation-models}.

\subsection{Implications for chemical models and $\letCO{X}(Z)$}
\label{sec:discussion:implications}
\begin{figure*}
    \centering
    \includegraphics[width=\linewidth]{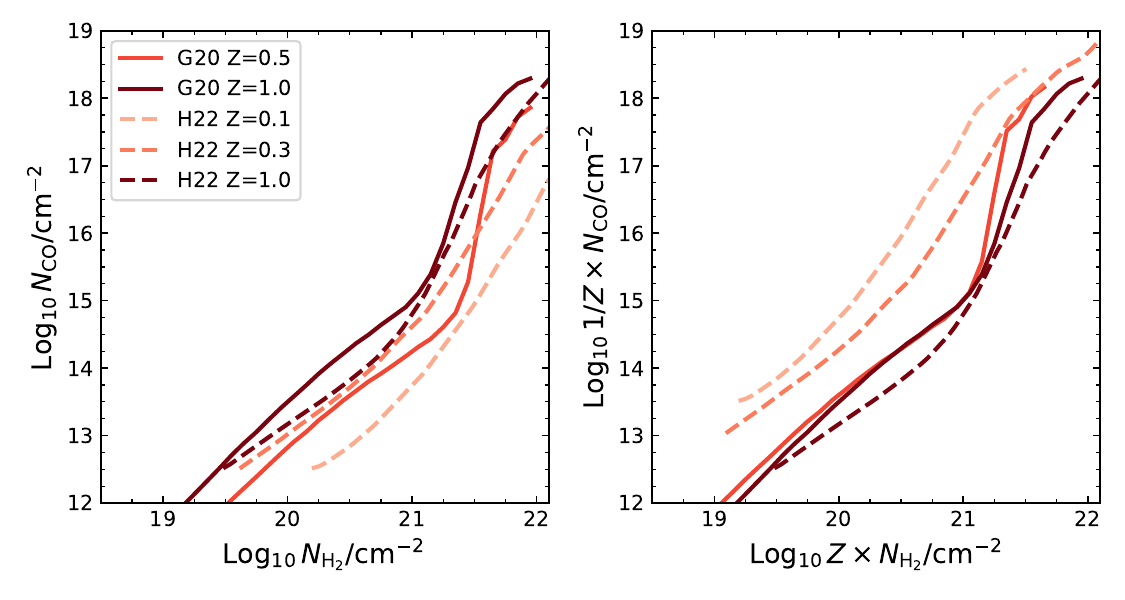}
    \caption{Actual (left) and abundance-rescaled (right) \isoCO{12}\ and \molH\ column densities at different metallicities predicted from two simulations (\citealt{Gong:2020-TIGRESS-XCO-variation}, G20, solid lines; \citealt{Hu:2022-CO-Z}, H22, dashed lines). The left panel shows the median predicted \NCO\ as a function of \NmolH in simulations with different metallicities. The right panel ``reverses'' a hypothetical effect of metallicity on the \NCO(\NmolH) relation. \NCO\ is assumed to shift ``vertically'' due to the decrease in carbon abundance (here assumed to be proportional to metallicity $Z$) and to shift ``horizontally'' due to the smaller dust optical depth $A_V$ per hydrogen nucleus. Reversing these two effects lines up the G20 trends but overcorrects the H22 trends. \autoref{fig:H2-CO-data} shows that this reversal also overcorrects the actual measurements.}
    \label{fig:NH2-NCO-models}
\end{figure*}

In \S \ref{sec:results}, we show that evolution of \NCO(\NmolH) does not appear to be consistent with a scenario where the \NmolH\ required to reach a particular $\NCO/(\alpha_C\times\NmolH)$ scales with the dust-to-gas ratio.
We showed this quantitatively in \S \ref{sec:results-NCO-NH2} but first demonstrated it visually at the beginning of \S \ref{sec:results} by ``reversing'' the expected shifts due to changes in the carbon abundance and dust-to-gas ratio.
We can do the same exercise with predictions for \NCO(\NmolH) from simulations.

The left panel of \autoref{fig:NH2-NCO-models} shows median \NCO(\NmolH) relations based on the work of \citet[G20]{Gong:2020-TIGRESS-XCO-variation} and taken from Figure 5 of \citet[H22]{Hu:2022-CO-Z}.
We calculated the G20 relations using the publicly released TIGRESS simulation outputs.
\citet{Gong:2020-TIGRESS-XCO-variation} post-process a hydrodynamical simulation to calculate \molH\ and CO abundances assuming chemical equilibrium.
\citet{Hu:2022-CO-Z} run time-dependent \molH\ chemistry during a hydrodynamical simulation and post-process the output to get CO abundances assuming chemical equilibrium.

The right panel shows the result of reversing the abundance and dust-to-gas ratio shifts.
The H22 relations are clearly over-corrected while the the G20 relations line up almost perfectly for \NmolH\ less than $10^{21}$ \cmmt\ and are close at higher columns.
The break in the G20 relations happens at the same ``corrected'' \NmolH.
The location of the break is thought to be the point where CO becomes sufficiently shielded by dust against photodissociation, so it makes sense that this location would scale with the dust-to-gas ratio.

Under the ``correction'' procedure, the measurements behave like the H22 predictions.
The strong interpretation of this behavior is \molH\ is not in chemical equilibrium in the ISM.
One consequence of this would be an upper bound on the dynamical time of gas in molecular clouds: for \molH\ to be out of equilibrium, the dynamical time has to be shorter than the \molH\ formation timescale.
A more practical consequence would be that analyses of CO emission should be done using outputs of time-dependent calculations rather than using equilibrium models.
Finally, the agreement with this particular time-dependent calculation would imply that the dependence of \letCO{X}\ on metallicity is slower than linear.

To sound a few notes of caution, the comparison done here was visual rather than quantitative; we examined only one set of predictions of each kind; and the behavior predicted by \citet{Hu:2022-CO-Z} is not universally found in simulations with time-dependent chemistry.
We intend to remedy the first two issues in later work.
An example of the third issue is \citet{Richings:2016-non-eq-CO-mol-clouds}, who find that non-equilibrium chemistry can suppress CO more than \molH.
Understanding differences between non-equilibrium chemical models will require a wider and more thorough set of comparisons.

\subsection{Regarding \citet{Balashev:2025aa}}
Shortly before we submitted this work for publication, we became aware of a similar analysis of CO absorption in the Magellanic Clouds, \citet{Balashev:2025aa}. 
There are several differences in choice of subject matter between the articles.
We include lines of sight with lower \molH\ column densities (though these are all non-detections), describe a different and novel procedure for disentangling ISM absorption from the stellar continuum, and combine our results with Milky Way CO measurements and simulation outputs from the literature to estimate how the CO-dark gas fraction changes with metallicity. 
\citet{Balashev:2025aa} include sightline-specific models of collisional and radiative CO excitation and provide more context on high redshift systems. 

Our measurements are statistically consistent with theirs with two exceptions.
For their three detections, we find a significantly lower (0.8 dex) CO column density toward AV 456 (Sk 143 in their work) and higher though consistent CO column densities toward \sknum{68}{137}\ and \sknum{67}{2}. 
We classify five of their non-detections as detections.
For four of these cases, our column density measurements are consistent with their upper limits. 
Our measurement for AV 476 is at least half a dex higher than their upper limit. 
We hypothesize that the differences in column densities are related to differences in the treatment of the stellar continuum.

\section{Conclusion} \label{sec:conclusion}
We use ultraviolet absorption spectroscopy to measure carbon monoxide column densities \NCO\ along 50 lines of sight through the Large and Small Magellanic Clouds (LMC and SMC).
We combine our new \NCO\ measurements with archival molecular hydrogen column densities \NmolH\ from \citet{Welty:2012vl} in order to study the relation between \NmolH\ and \NCO\ at 1/2 (LMC) and 1/5 (SMC) solar metallicity.
The way in which this relation changes with metallicity is an informative constraint for models of interstellar medium chemistry.
This relation encodes the relative ``sizes'' of the \molH-rich and CO-rich parts of a molecular cloud.
The evolution of these relative sizes is one of the main drivers for the metallicity dependence of the CO emission to molecular gas mass conversion factor, $\letCO{X}$.

The \NmolH\ at which the fraction of carbon in CO reaches a particular value can be used as a proxy for these relative sizes.
Comparing our LMC and SMC measurements with literature values for the Milky Way, we find that this threshold \NmolH\ changes by about 0.05 dex from the Milky Way to the LMC and about 0.1 dex from the Milky Way to the SMC (\S\ref{sec:results-NCO-NH2} and \autoref{fig:fit-NCO-NH2}). 
These values are smaller than expected from chemical equilibrium calculations, where the shift is roughly proportional to the change in the gas-to-dust ratio.
We argue in \S\ref{sec:discussion:implications} that these relatively small shifts are tentative evidence for a scenario where the \molH\ abundance does not have time to reach equilibrium while the CO abundance does. 

Estimating $\letCO{X}$ from our measurements requires a model for gas excitation, which is beyond the scope of the current work.
By combining our measurements of the relation between \NmolH\ and \NCO\ with a simulated \NmolH\ distribution from the literature, we are however able to estimate the fraction of \molH\ mass associated with detectable CO emission. 
We find that the CO-bright \molH\ mass fraction in the LMC and SMC is about 0.75 and 0.55, respectively, of the Milky Way CO-bright fraction. 
The decline in the CO-bright fraction is less steep than the decline in metallicity, which is qualitative evidence for a relatively shallow dependence of $\letCO{X}$ on metallicity.
In terms of some recent estimates and predictions of the power law index $\alpha$ assuming $\letCO{X}\propto Z^\alpha$, our measurements are more consistent with $\alpha \gtrsim -1$ \citep{Gong:2020-TIGRESS-XCO-variation,Hu:2022-CO-Z} or $\alpha \approx -1.6$ \citep{Accurso:2017aa-XCO} than $\alpha \approx -2.4$ \citep{Schruba:2012} or -3.4 \citep{Madden:2020aa-XCO}.

\appendix
\section{Upper limits}
\tablewidth{\linewidth}
\begin{deluxetable}{lc}
\tablecaption{Column density upper limits in the LMC \label{tab:LMC-uplims}}
\tablehead{
ULLYSES ID & \colhead{Log$_{10}$\NCO} \\
 & [\cmmt]
}
\startdata
SK-71D8	& 12.83 \\
SK-69D220 & 13.59\\
SK-67D105	& 13.43 \\
PGMW-3070	& 13.09 \\
SK-66D35	& 12.67 \\
SK-67D5	& 12.88 \\
SK-68D52	& 13.16 \\
BI-184	& 13.27 \\
SK-69D246	& 13.46 \\
VFTS-72	& 13.72 \\
SK-68D135	& 13.68 \\
SK-70D115	& 12.56 \\
SK-68D155	& 13.51 \\
BI-237	& 12.59 \\
SK-69D224	& 12.22 \\
SK-71D50	& 12.89 \\
SK-66D19	& 13.48 \\
SK-68D129	& 13.63 \\
SK-71D46	& 13.08 \\
SK-69D279	& 12.77 \\
SK-68D26	& 12.79 \\
\enddata
\tablecomments{Upper limits on \isoCO{12}\ column densities where CO is not detected at a significant level. Limits are 95th percentiles of the posterior probability distribution over the column density.}
\end{deluxetable}
\begin{deluxetable}{lc}
\tablecaption{Column density upper limits in the SMC \label{tab:SMC-uplims}}
\tablehead{
ULLYSES ID & \colhead{Log$_{10}$\NCO} \\
 & [\cmmt]
}
\startdata
AV-6 &	13.04 \\
AV-479 &	13.07 \\
AV-104 &	13.70 \\
AV-261 &	13.08 \\
AV-210 &	13.57 \\
AV-207 &	13.14 \\
AV-388 &	12.30 \\
AV-95 &	12.73 \\
2DFS-999 &	13.91 \\
AV-304 &    13.41 \\
AV-215 &	13.09 \\
AV-170 &	13.03 \\
AV-490 &	12.37 \\
AV-175 &	12.83 \\
AV-435 &	12.95 \\
AV-80 &	12.65 \\
AV-16 &	13.33 \\
AV-472 &	13.12 \\
AV-18 &	12.44 \\
NGC346-ELS-026 &	12.53 \\
AV-26 &	13.09 \\
\enddata
\tablecomments{Upper limits on \isoCO{12}\ column densities where CO is not detected at a significant level. Limits are 95th percentiles of the posterior probability distribution over the column density.}
\end{deluxetable}
Tables \ref{tab:LMC-uplims} and \ref{tab:SMC-uplims} list upper limits on the column density of CO toward targets without a statistically significant detection of CO absorption.
The limits are 95th percentiles of the probability distribution over $\lten$ CO column density.

\begin{acknowledgements}
We thank the anonymous referee for their report, which has improved this paper. This work was done with support from program \#HST-AR-16635 provided by NASA through a grant from the Space Telescope Science Institute.
\end{acknowledgements}

\software{\texttt{astropy} \citep{Astropy-Collaboration:2013uv,Astropy-Collaboration:2018vm},
\\ \texttt{numpyro} \citep{Phan:2019-numpyro},
\\ \texttt{tinygp} \citep{Foreman-Mackey:2024-tinygp},
\\ \texttt{linetools} \citep{Prochaska:2017vh},
\\ \texttt{matplotlib} \citep{Hunter:2007ux},
\\ \texttt{numpy} \citep{Harris:2020ti},
\\
\texttt{pandas} \citep{McKinney:2010vw}
}

\bibliography{main}

\begin{thebibliography}{}
\expandafter\ifx\csname natexlab\endcsname\relax\def\natexlab#1{#1}\fi
\providecommand{\url}[1]{\href{#1}{#1}}

\bibitem[{{Accurso} {et~al.}(2017){Accurso}, {Saintonge}, {Catinella},
  {Cortese}, {Dav{\'e}}, {Dunsheath}, {Genzel}, {Gracia-Carpio}, {Heckman},
  {Jimmy}, {Kramer}, {Li}, {Lutz}, {Schiminovich}, {Schuster}, {Sternberg},
  {Sturm}, {Tacconi}, {Tran}, \& {Wang}}]{Accurso:2017aa-XCO}
{Accurso}, G., {Saintonge}, A., {Catinella}, B., {et~al.} 2017, \mnras, 470,
  4750

\bibitem[{{Andr{\'e}} {et~al.}(2004){Andr{\'e}}, {Le Petit}, {Sonnentrucker},
  {Ferlet}, {Roueff}, {Civeit}, {D{\'e}sert}, {Lacour}, \&
  {Vidal-Madjar}}]{Andre:2004aa}
{Andr{\'e}}, M.~K., {Le Petit}, F., {Sonnentrucker}, P., {et~al.} 2004, \aap,
  422, 483

\bibitem[{{Astropy Collaboration} {et~al.}(2013){Astropy Collaboration},
  {Robitaille}, {Tollerud}, {Greenfield}, {Droettboom}, {Bray}, {Aldcroft},
  {Davis}, {Ginsburg}, {Price-Whelan}, {Kerzendorf}, {Conley}, {Crighton},
  {Barbary}, {Muna}, {Ferguson}, {Grollier}, {Parikh}, {Nair}, {Unther},
  {Deil}, {Woillez}, {Conseil}, {Kramer}, {Turner}, {Singer}, {Fox}, {Weaver},
  {Zabalza}, {Edwards}, {Azalee Bostroem}, {Burke}, {Casey}, {Crawford},
  {Dencheva}, {Ely}, {Jenness}, {Labrie}, {Lim}, {Pierfederici}, {Pontzen},
  {Ptak}, {Refsdal}, {Servillat}, \&
  {Streicher}}]{Astropy-Collaboration:2013uv}
{Astropy Collaboration}, {Robitaille}, T.~P., {Tollerud}, E.~J., {et~al.} 2013,
  \aap, 558, A33

\bibitem[{{Astropy Collaboration} {et~al.}(2018){Astropy Collaboration},
  {Price-Whelan}, {Sip{\H{o}}cz}, {G{\"u}nther}, {Lim}, {Crawford}, {Conseil},
  {Shupe}, {Craig}, {Dencheva}, {Ginsburg}, {Vand erPlas}, {Bradley},
  {P{\'e}rez-Su{\'a}rez}, {de Val-Borro}, {Aldcroft}, {Cruz}, {Robitaille},
  {Tollerud}, {Ardelean}, {Babej}, {Bach}, {Bachetti}, {Bakanov}, {Bamford},
  {Barentsen}, {Barmby}, {Baumbach}, {Berry}, {Biscani}, {Boquien}, {Bostroem},
  {Bouma}, {Brammer}, {Bray}, {Breytenbach}, {Buddelmeijer}, {Burke},
  {Calderone}, {Cano Rodr{\'\i}guez}, {Cara}, {Cardoso}, {Cheedella}, {Copin},
  {Corrales}, {Crichton}, {D'Avella}, {Deil}, {Depagne}, {Dietrich}, {Donath},
  {Droettboom}, {Earl}, {Erben}, {Fabbro}, {Ferreira}, {Finethy}, {Fox},
  {Garrison}, {Gibbons}, {Goldstein}, {Gommers}, {Greco}, {Greenfield},
  {Groener}, {Grollier}, {Hagen}, {Hirst}, {Homeier}, {Horton}, {Hosseinzadeh},
  {Hu}, {Hunkeler}, {Ivezi{\'c}}, {Jain}, {Jenness}, {Kanarek}, {Kendrew},
  {Kern}, {Kerzendorf}, {Khvalko}, {King}, {Kirkby}, {Kulkarni}, {Kumar},
  {Lee}, {Lenz}, {Littlefair}, {Ma}, {Macleod}, {Mastropietro}, {McCully},
  {Montagnac}, {Morris}, {Mueller}, {Mumford}, {Muna}, {Murphy}, {Nelson},
  {Nguyen}, {Ninan}, {N{\"o}the}, {Ogaz}, {Oh}, {Parejko}, {Parley}, {Pascual},
  {Patil}, {Patil}, {Plunkett}, {Prochaska}, {Rastogi}, {Reddy Janga},
  {Sabater}, {Sakurikar}, {Seifert}, {Sherbert}, {Sherwood-Taylor}, {Shih},
  {Sick}, {Silbiger}, {Singanamalla}, {Singer}, {Sladen}, {Sooley},
  {Sornarajah}, {Streicher}, {Teuben}, {Thomas}, {Tremblay}, {Turner},
  {Terr{\'o}n}, {van Kerkwijk}, {de la Vega}, {Watkins}, {Weaver}, {Whitmore},
  {Woillez}, {Zabalza}, \& {Astropy
  Contributors}}]{Astropy-Collaboration:2018vm}
{Astropy Collaboration}, {Price-Whelan}, A.~M., {Sip{\H{o}}cz}, B.~M., {et~al.}
  2018, \aj, 156, 123

\bibitem[{{Balashev} {et~al.}(2025){Balashev}, {Kosenko}, \&
  {Noterdaeme}}]{Balashev:2025aa}
{Balashev}, S., {Kosenko}, D., \& {Noterdaeme}, P. 2025, arXiv e-prints,
  arXiv:2503.12516

\bibitem[{{Balashev} {et~al.}(2017){Balashev}, {Noterdaeme}, {Rahmani},
  {Klimenko}, {Ledoux}, {Petitjean}, {Srianand}, {Ivanchik}, \&
  {Varshalovich}}]{Balashev:2017-CO-dark-DLA}
{Balashev}, S.~A., {Noterdaeme}, P., {Rahmani}, H., {et~al.} 2017, \mnras, 470,
  2890

\bibitem[{{Bally} \& {Langer}(1982)}]{Bally:1982-CO-isotope-photodestruction}
{Bally}, J., \& {Langer}, W.~D. 1982, \apj, 255, 143

\bibitem[{{Bialy} \& {Sternberg}(2015)}]{Bialy:2015}
{Bialy}, S., \& {Sternberg}, A. 2015, \mnras, 450, 4424

\bibitem[{{Bisbas} {et~al.}(2015){Bisbas}, {Papadopoulos}, \&
  {Viti}}]{Bisbas:2015}
{Bisbas}, T.~G., {Papadopoulos}, P.~P., \& {Viti}, S. 2015, \apj, 803, 37

\bibitem[{{Bisbas} {et~al.}(2017){Bisbas}, {van Dishoeck}, {Papadopoulos},
  {Sz{\H{u}}cs}, {Bialy}, \& {Zhang}}]{Bisbas:2017}
{Bisbas}, T.~G., {van Dishoeck}, E.~F., {Papadopoulos}, P.~P., {et~al.} 2017,
  \apj, 839, 90

\bibitem[{{Bluhm} \& {de Boer}(2001)}]{Bluhm:2001aa-CO-Sk-69-246}
{Bluhm}, H., \& {de Boer}, K.~S. 2001, \aap, 379, 82

\bibitem[{{Burgh} {et~al.}(2010){Burgh}, {France}, \& {Jenkins}}]{Burgh:2010wd}
{Burgh}, E.~B., {France}, K., \& {Jenkins}, E.~B. 2010, \apj, 708, 334

\bibitem[{{Burgh} {et~al.}(2007){Burgh}, {France}, \&
  {McCandliss}}]{Burgh:2007vc}
{Burgh}, E.~B., {France}, K., \& {McCandliss}, S.~R. 2007, \apj, 658, 446

\bibitem[{{Chin} {et~al.}(1999){Chin}, {Henkel}, {Langer}, \&
  {Mauersberger}}]{Chin:1999-LMC-12C-13C}
{Chin}, Y.-n., {Henkel}, C., {Langer}, N., \& {Mauersberger}, R. 1999, \apjl,
  512, L143

\bibitem[{{Clark} {et~al.}(2023){Clark}, {Roman-Duval}, {Gordon}, {Bot},
  {Smith}, \& {Hagen}}]{Clark:2023-DGR-Z-Sigma-variation}
{Clark}, C. J.~R., {Roman-Duval}, J.~C., {Gordon}, K.~D., {et~al.} 2023, \apj,
  946, 42

\bibitem[{{Crenny} \& {Federman}(2004)}]{Crenny:2004ul}
{Crenny}, T., \& {Federman}, S.~R. 2004, \apj, 605, 278

\bibitem[{{Dapr{\`a}} {et~al.}(2016){Dapr{\`a}}, {Niu}, {Salumbides}, {Murphy},
  \& {Ubachs}}]{Dapra:2016:CO}
{Dapr{\`a}}, M., {Niu}, M.~L., {Salumbides}, E.~J., {Murphy}, M.~T., \&
  {Ubachs}, W. 2016, \apj, 826, 192

\bibitem[{{Eidelsberg} \&
  {Rostas}(2003)}]{Eidelsberg:2003-intersystem-CO-transitions}
{Eidelsberg}, M., \& {Rostas}, F. 2003, \apjs, 145, 89

\bibitem[{{Foreman-Mackey} {et~al.}(2024){Foreman-Mackey}, {Yu}, {Yadav},
  {Reynolds Becker}, {Caplar}, {Huppenkothen}, {Killestein}, {Tronsgaard},
  {Rashid}, \& {Schmerler}}]{Foreman-Mackey:2024-tinygp}
{Foreman-Mackey}, D., {Yu}, W., {Yadav}, S., {et~al.} 2024, {dfm/tinygp: The
  tiniest of Gaussian Process libraries}, vv0.3.0,  Zenodo,
  doi:10.5281/zenodo.10463641

\bibitem[{{Gerin} \& {Liszt}(2021)}]{Gerin:2021-COplus-CO}
{Gerin}, M., \& {Liszt}, H. 2021, \aap, 648, A38

\bibitem[{{Gillmon} {et~al.}(2006){Gillmon}, {Shull}, {Tumlinson}, \&
  {Danforth}}]{Gillmon:2006tl}
{Gillmon}, K., {Shull}, J.~M., {Tumlinson}, J., \& {Danforth}, C. 2006, \apj,
  636, 891

\bibitem[{{Gong} {et~al.}(2018){Gong}, {Ostriker}, \&
  {Kim}}]{Gong:2018-TIGRESS-XCO-fiducial}
{Gong}, M., {Ostriker}, E.~C., \& {Kim}, C.-G. 2018, \apj, 858, 16

\bibitem[{{Gong} {et~al.}(2020){Gong}, {Ostriker}, {Kim}, \&
  {Kim}}]{Gong:2020-TIGRESS-XCO-variation}
{Gong}, M., {Ostriker}, E.~C., {Kim}, C.-G., \& {Kim}, J.-G. 2020, \apj, 903,
  142

\bibitem[{{Green} {et~al.}(2012){Green}, {Froning}, {Osterman}, {Ebbets},
  {Heap}, {Leitherer}, {Linsky}, {Savage}, {Sembach}, {Shull}, {Siegmund},
  {Snow}, {Spencer}, {Stern}, {Stocke}, {Welsh}, {B{\'e}land}, {Burgh},
  {Danforth}, {France}, {Keeney}, {McPhate}, {Penton}, {Andrews},
  {Brownsberger}, {Morse}, \& {Wilkinson}}]{Green_2012_COS_citation}
{Green}, J.~C., {Froning}, C.~S., {Osterman}, S., {et~al.} 2012, \apj, 744, 60

\bibitem[{{Grishunin} {et~al.}(2023){Grishunin}, {Weiss}, {Colombo},
  {Chevance}, {Chen}, {G{\"u}sten}, {Rubio}, {Hunt}, {Wyrowski}, {Harrington},
  {Menten}, \& {Herrera-Camus}}]{Grishunin:2023-APEX-LMC-initial}
{Grishunin}, K., {Weiss}, A., {Colombo}, D., {et~al.} 2023, arXiv e-prints,
  arXiv:2310.20701

\bibitem[{{Hainich} {et~al.}(2019){Hainich}, {Ramachandran}, {Shenar},
  {Sander}, {Todt}, {Gruner}, {Oskinova}, \& {Hamann}}]{Hainich_2019_PoWR_OB}
{Hainich}, R., {Ramachandran}, V., {Shenar}, T., {et~al.} 2019, \aap, 621, A85

\bibitem[{Harris {et~al.}(2020)Harris, Millman, van~der Walt, Gommers,
  Virtanen, Cournapeau, Wieser, Taylor, Berg, Smith, Kern, Picus, Hoyer, van
  Kerkwijk, Brett, Haldane, del R{\'{i}}o, Wiebe, Peterson,
  G{\'{e}}rard-Marchant, Sheppard, Reddy, Weckesser, Abbasi, Gohlke, \&
  Oliphant}]{Harris:2020ti}
Harris, C.~R., Millman, K.~J., van~der Walt, S.~J., {et~al.} 2020, Nature, 585,
  357.
\newblock \url{https://doi.org/10.1038/s41586-020-2649-2}

\bibitem[{{Hu} {et~al.}(2022){Hu}, {Schruba}, {Sternberg}, \& {van
  Dishoeck}}]{Hu:2022-CO-Z}
{Hu}, C.-Y., {Schruba}, A., {Sternberg}, A., \& {van Dishoeck}, E.~F. 2022,
  \apj, 931, 28

\bibitem[{{Hu} {et~al.}(2021){Hu}, {Sternberg}, \& {van Dishoeck}}]{Hu:2021uj}
{Hu}, C.-Y., {Sternberg}, A., \& {van Dishoeck}, E.~F. 2021, \apj, 920, 44

\bibitem[{{Hu} {et~al.}(2023){Hu}, {Sternberg}, \& {van
  Dishoeck}}]{Hu:2023-lowZ-dust-evo}
---. 2023, \apj, 952, 140

\bibitem[{{Hunter}(2007)}]{Hunter:2007ux}
{Hunter}, J.~D. 2007, Computing in Science and Engineering, 9, 90

\bibitem[{{Kass} \& Adrian(1995)}]{Kass:1995vb}
{Kass}, E.~R., \& Adrian, R.~E. 1995, Journal of the American Statistical
  Association, 90, 773

\bibitem[{{Kim} \& {Ostriker}(2017)}]{Kim:2017-TIGRESS-general-ref}
{Kim}, C.-G., \& {Ostriker}, E.~C. 2017, \apj, 846, 133

\bibitem[{{Kimble} {et~al.}(1998){Kimble}, {Woodgate}, {Bowers}, {Kraemer},
  {Kaiser}, {Gull}, {Heap}, {Danks}, {Boggess}, {Green}, {Hutchings},
  {Jenkins}, {Joseph}, {Linsky}, {Maran}, {Moos}, {Roesler}, {Timothy},
  {Weistrop}, {Grady}, {Loiacono}, {Brown}, {Brumfield}, {Content}, {Feinberg},
  {Isaacs}, {Krebs}, {Krueger}, {Melcher}, {Rebar}, {Vitagliano}, {Yagelowich},
  {Meyer}, {Hood}, {Argabright}, {Becker}, {Bottema}, {Breyer}, {Bybee},
  {Christon}, {Delamere}, {Dorn}, {Downey}, {Driggers}, {Ebbets}, {Gallegos},
  {Garner}, {Hetlinger}, {Lettieri}, {Ludtke}, {Michika}, {Nyquist}, {Rose},
  {Stocker}, {Sullivan}, {Van Houten}, {Woodruff}, {Baum}, {Hartig}, {Balzano},
  {Biagetti}, {Blades}, {Bohlin}, {Clampin}, {Doxsey}, {Ferguson},
  {Goudfrooij}, {Hulbert}, {Kutina}, {McGrath}, {Lindler}, {Beck}, {Feggans},
  {Plait}, {Sandoval}, {Hill}, {Collins}, {Cornett}, {Fowler}, {Hill},
  {Landsman}, {Malumuth}, {Standley}, {Blouke}, {Grusczak}, {Reed}, {Robinson},
  {Valenti}, \& {Wolfe}}]{Kimble:1998_STIS_performance}
{Kimble}, R.~A., {Woodgate}, B.~E., {Bowers}, C.~W., {et~al.} 1998, \apjl, 492,
  L83

\bibitem[{{Koenigsberger} {et~al.}(2001){Koenigsberger}, {Georgiev},
  {Peimbert}, {Walborn}, {Barb{\'a}}, {Niemela}, {Morrell}, {Tsvetanov}, \&
  {Schulte-Ladbeck}}]{Koenigsberger:2001-implausible-CO-detection}
{Koenigsberger}, G., {Georgiev}, L., {Peimbert}, M., {et~al.} 2001, \aj, 121,
  267

\bibitem[{{Lennon} {et~al.}(2017){Lennon}, {van der Marel}, {Ramos Lerate},
  {O'Mullane}, \& {Sahlmann}}]{Lennon:2017-sk-67-2-hypervelocity}
{Lennon}, D.~J., {van der Marel}, R.~P., {Ramos Lerate}, M., {O'Mullane}, W.,
  \& {Sahlmann}, J. 2017, \aap, 603, A75

\bibitem[{{Liszt}(2017)}]{Liszt:2017-CO-fractionation}
{Liszt}, H.~S. 2017, \apj, 835, 138

\bibitem[{{Madden} {et~al.}(2020){Madden}, {Cormier}, {Hony}, {Lebouteiller},
  {Abel}, {Galametz}, {De Looze}, {Chevance}, {Polles}, {Lee}, {Galliano},
  {Lambert-Huyghe}, {Hu}, \& {Ramambason}}]{Madden:2020aa-XCO}
{Madden}, S.~C., {Cormier}, D., {Hony}, S., {et~al.} 2020, \aap, 643, A141

\bibitem[{{Morton} \& {Noreau}(1994)}]{Morton:1994ty}
{Morton}, D.~C., \& {Noreau}, L. 1994, \apjs, 95, 301

\bibitem[{{Phan} {et~al.}(2019){Phan}, {Pradhan}, \&
  {Jankowiak}}]{Phan:2019-numpyro}
{Phan}, D., {Pradhan}, N., \& {Jankowiak}, M. 2019, arXiv e-prints,
  arXiv:1912.11554

\bibitem[{{Prochaska} {et~al.}(2017){Prochaska}, {Tejos}, {Crighton},
  {jnburchett}, {tiffanyhsyu}, {Tuo-Ji}, {marijana777}, {ktirimba}, {jhennawi},
  {Cooke}, {O'Meara}, \& {Werk}}]{Prochaska:2017vh}
{Prochaska}, J.~X., {Tejos}, N., {Crighton}, N., {et~al.} 2017,
  {Linetools/Linetools: Third Minor Release}, vv0.3,  Zenodo,
  doi:10.5281/zenodo.1036773

\bibitem[{{Rachford} {et~al.}(2002){Rachford}, {Snow}, {Tumlinson}, {Shull},
  {Blair}, {Ferlet}, {Friedman}, {Gry}, {Jenkins}, {Morton}, {Savage},
  {Sonnentrucker}, {Vidal-Madjar}, {Welty}, \& {York}}]{Rachford:2002wo}
{Rachford}, B.~L., {Snow}, T.~P., {Tumlinson}, J., {et~al.} 2002, \apj, 577,
  221

\bibitem[{{Ramambason} {et~al.}(2024){Ramambason}, {Lebouteiller}, {Madden},
  {Galliano}, {Richardson}, {Saintonge}, {De Looze}, {Chevance}, {Abel},
  {Hernandez}, \& {Braine}}]{Ramambason:2024-XCO-split}
{Ramambason}, L., {Lebouteiller}, V., {Madden}, S.~C., {et~al.} 2024, \aap,
  681, A14

\bibitem[{{Ranjan} {et~al.}(2018){Ranjan}, {Noterdaeme}, {Krogager},
  {Petitjean}, {Balashev}, {Bialy}, {Srianand}, {Gupta}, {Fynbo}, {Ledoux}, \&
  {Laursen}}]{Ranjan:2018-strong-H2-absorber}
{Ranjan}, A., {Noterdaeme}, P., {Krogager}, J.~K., {et~al.} 2018, \aap, 618,
  A184

\bibitem[{{Richings} \& {Schaye}(2016)}]{Richings:2016-non-eq-CO-mol-clouds}
{Richings}, A.~J., \& {Schaye}, J. 2016, \mnras, 460, 2297

\bibitem[{{Roman-Duval} {et~al.}(2017){Roman-Duval}, {Bot}, {Chastenet}, \&
  {Gordon}}]{Roman-Duval:2017aa}
{Roman-Duval}, J., {Bot}, C., {Chastenet}, J., \& {Gordon}, K. 2017, \apj, 841,
  72

\bibitem[{{Roman-Duval} {et~al.}(2014){Roman-Duval}, {Gordon}, {Meixner},
  {Bot}, {Bolatto}, {Hughes}, {Wong}, {Babler}, {Bernard}, {Clayton}, {Fukui},
  {Galametz}, {Galliano}, {Glover}, {Hony}, {Israel}, {Jameson},
  {Lebouteiller}, {Lee}, {Li}, {Madden}, {Misselt}, {Montiel}, {Okumura},
  {Onishi}, {Panuzzo}, {Reach}, {Remy-Ruyer}, {Robitaille}, {Rubio}, {Sauvage},
  {Seale}, {Sewilo}, {Staveley-Smith}, \& {Zhukovska}}]{Roman-Duval:2014aa}
{Roman-Duval}, J., {Gordon}, K.~D., {Meixner}, M., {et~al.} 2014, \apj, 797, 86

\bibitem[{{Roman-Duval} {et~al.}(2020){Roman-Duval}, {Proffitt}, {Taylor},
  {Monroe}, {Fischer}, {Fischer}, {Fullerton}, {Aloisi}, {Britt}, {Busko},
  {Carlberg}, {De Rosa}, {Jedrzejewski}, {Lockwood}, {Frazer}, {Hernandez},
  {James}, {Oliveira}, {Plesha}, {Riedel}, {Riley}, {Sahnow}, {Sankrit},
  {Shaw}, {Smith}, {Sohn}, {Som}, {Ubeda}, \&
  {Welty}}]{Roman-Duval:2020aa-ULLYSES-citation}
{Roman-Duval}, J., {Proffitt}, C.~R., {Taylor}, J.~M., {et~al.} 2020, Research
  Notes of the American Astronomical Society, 4, 205

\bibitem[{{Roman-Duval} {et~al.}(2022){Roman-Duval}, {Jenkins}, {Tchernyshyov},
  {Clark}, {De Cia}, {Gordon}, {Hamanowicz}, {Lebouteiller}, {Rafelski},
  {Sandstrom}, {Werk}, \& {Merica-Jones}}]{Roman-Duval:2022aa}
{Roman-Duval}, J., {Jenkins}, E.~B., {Tchernyshyov}, K., {et~al.} 2022, \apj,
  928, 90

\bibitem[{{Roueff} {et~al.}(2015){Roueff}, {Loison}, \&
  {Hickson}}]{Roueff:2015-fractionation-reactions}
{Roueff}, E., {Loison}, J.~C., \& {Hickson}, K.~M. 2015, \aap, 576, A99

\bibitem[{{Russell} \& {Dopita}(1992)}]{Russell:1992aa}
{Russell}, S.~C., \& {Dopita}, M.~A. 1992, \apj, 384, 508

\bibitem[{{Sander} {et~al.}(2012){Sander}, {Hamann}, \&
  {Todt}}]{Sander_2012_PoWR_WC}
{Sander}, A., {Hamann}, W.~R., \& {Todt}, H. 2012, \aap, 540, A144

\bibitem[{{Schruba} {et~al.}(2012){Schruba}, {Leroy}, {Walter}, {Bigiel},
  {Brinks}, {de Blok}, {Kramer}, {Rosolowsky}, {Sandstrom}, {Schuster},
  {Usero}, {Weiss}, \& {Wiesemeyer}}]{Schruba:2012}
{Schruba}, A., {Leroy}, A.~K., {Walter}, F., {et~al.} 2012, \aj, 143, 138

\bibitem[{{Sheffer} {et~al.}(2008){Sheffer}, {Rogers}, {Federman}, {Abel},
  {Gredel}, {Lambert}, \& {Shaw}}]{Sheffer:2008vc}
{Sheffer}, Y., {Rogers}, M., {Federman}, S.~R., {et~al.} 2008, \apj, 687, 1075

\bibitem[{{Sheffer} {et~al.}(2007){Sheffer}, {Rogers}, {Federman}, {Lambert},
  \& {Gredel}}]{Sheffer:2007-12CO-13CO}
{Sheffer}, Y., {Rogers}, M., {Federman}, S.~R., {Lambert}, D.~L., \& {Gredel},
  R. 2007, \apj, 667, 1002

\bibitem[{{Sonnentrucker} {et~al.}(2007){Sonnentrucker}, {Welty}, {Thorburn},
  \& {York}}]{Sonnentrucker:2007ux}
{Sonnentrucker}, P., {Welty}, D.~E., {Thorburn}, J.~A., \& {York}, D.~G. 2007,
  \apjs, 168, 58

\bibitem[{{Sz{\H{u}}cs} {et~al.}(2014){Sz{\H{u}}cs}, {Glover}, \&
  {Klessen}}]{Szucs:2014-CO-fractionation}
{Sz{\H{u}}cs}, L., {Glover}, S. C.~O., \& {Klessen}, R.~S. 2014, \mnras, 445,
  4055

\bibitem[{{Tchernyshyov}(2020)}]{Tchernyshyov:2020-amlc}
{Tchernyshyov}, K. 2020, \aj, 159, 64

\bibitem[{{Thompson} {et~al.}(2024){Thompson}, {Richings}, {Gibson},
  {Faucher-Gigu{\`e}re}, {Feldmann}, \& {Hayward}}]{Thompson:2024-neq-CO-H2}
{Thompson}, O.~A., {Richings}, A.~J., {Gibson}, B.~K., {et~al.} 2024, \mnras,
  532, 1948

\bibitem[{{Todt} {et~al.}(2015){Todt}, {Sander}, {Hainich}, {Hamann}, {Quade},
  \& {Shenar}}]{Todt_2015_PoWR_WN}
{Todt}, H., {Sander}, A., {Hainich}, R., {et~al.} 2015, \aap, 579, A75

\bibitem[{{Tumlinson} {et~al.}(2002){Tumlinson}, {Shull}, {Rachford},
  {Browning}, {Snow}, {Fullerton}, {Jenkins}, {Savage}, {Crowther}, {Moos},
  {Sembach}, {Sonneborn}, \& {York}}]{Tumlinson:2002tz}
{Tumlinson}, J., {Shull}, J.~M., {Rachford}, B.~L., {et~al.} 2002, \apj, 566,
  857

\bibitem[{{Visser} {et~al.}(2009){Visser}, {van Dishoeck}, \&
  {Black}}]{Visser:2009-CO-isotopologue-model}
{Visser}, R., {van Dishoeck}, E.~F., \& {Black}, J.~H. 2009, \aap, 503, 323

\bibitem[{{Wang} {et~al.}(2009){Wang}, {Chin}, {Henkel}, {Whiteoak}, \&
  {Cunningham}}]{Wang:2009-LMC-12C-13C}
{Wang}, M., {Chin}, Y.~N., {Henkel}, C., {Whiteoak}, J.~B., \& {Cunningham}, M.
  2009, \apj, 690, 580

\bibitem[{{Warin} {et~al.}(1996){Warin}, {Benayoun}, \&
  {Viala}}]{Warin:1996-CO-fractionation-models}
{Warin}, S., {Benayoun}, J.~J., \& {Viala}, Y.~P. 1996, \aap, 308, 535

\bibitem[{{Watson} {et~al.}(1976){Watson}, {Anicich}, \&
  {Huntress}}]{Watson:1976-CO-fractionation}
{Watson}, W.~D., {Anicich}, V.~G., \& {Huntress}, W.~T., J. 1976, \apjl, 205,
  L165

\bibitem[{{Welty} {et~al.}(2006){Welty}, {Federman}, {Gredel}, {Thorburn}, \&
  {Lambert}}]{Welty:2006MC-molecules}
{Welty}, D.~E., {Federman}, S.~R., {Gredel}, R., {Thorburn}, J.~A., \&
  {Lambert}, D.~L. 2006, \apjs, 165, 138

\bibitem[{{Welty} {et~al.}(2016){Welty}, {Lauroesch}, {Wong}, \&
  {York}}]{Welty:2016-MC-thermal-pressures}
{Welty}, D.~E., {Lauroesch}, J.~T., {Wong}, T., \& {York}, D.~G. 2016, \apj,
  821, 118

\bibitem[{{Welty} {et~al.}(2012){Welty}, {Xue}, \& {Wong}}]{Welty:2012vl}
{Welty}, D.~E., {Xue}, R., \& {Wong}, T. 2012, \apj, 745, 173

\bibitem[{{W}es {M}c{K}inney(2010)}]{McKinney:2010vw}
{W}es {M}c{K}inney. 2010, in {P}roceedings of the 9th {P}ython in {S}cience
  {C}onference, ed. {S}t\'efan van~der {W}alt \& {J}arrod {M}illman, 56 -- 61

\bibitem[{{Wolfire} {et~al.}(2010){Wolfire}, {Hollenbach}, \&
  {McKee}}]{Wolfire:2010-CO-dark-gas-layer}
{Wolfire}, M.~G., {Hollenbach}, D., \& {McKee}, C.~F. 2010, \apj, 716, 1191

\bibitem[{{Wong} {et~al.}(2017){Wong}, {Hughes}, {Tokuda}, {Indebetouw},
  {Bernard}, {Onishi}, {Wojciechowski}, {Bandurski}, {Kawamura}, {Roman-Duval},
  {Cao}, {Chen}, {Chu}, {Cui}, {Fukui}, {Montier}, {Muller}, {Ott}, {Paradis},
  {Pineda}, {Rosolowsky}, \& {Sewi{\l}o}}]{Wong:2017-MAGMA-DR3}
{Wong}, T., {Hughes}, A., {Tokuda}, K., {et~al.} 2017, \apj, 850, 139

\bibitem[{{Woodgate} {et~al.}(1998){Woodgate}, {Kimble}, {Bowers}, {Kraemer},
  {Kaiser}, {Danks}, {Grady}, {Loiacono}, {Brumfield}, {Feinberg}, {Gull},
  {Heap}, {Maran}, {Lindler}, {Hood}, {Meyer}, {Vanhouten}, {Argabright},
  {Franka}, {Bybee}, {Dorn}, {Bottema}, {Woodruff}, {Michika}, {Sullivan},
  {Hetlinger}, {Ludtke}, {Stocker}, {Delamere}, {Rose}, {Becker}, {Garner},
  {Timothy}, {Blouke}, {Joseph}, {Hartig}, {Green}, {Jenkins}, {Linsky},
  {Hutchings}, {Moos}, {Boggess}, {Roesler}, \&
  {Weistrop}}]{Woodgate:1998_STIS_design}
{Woodgate}, B.~E., {Kimble}, R.~A., {Bowers}, C.~W., {et~al.} 1998, \pasp, 110,
  1183

\bibitem[{{Zsarg{\'o}} \& {Federman}(2003)}]{Zsargo:2003-nonthermal-CHplus}
{Zsarg{\'o}}, J., \& {Federman}, S.~R. 2003, \apj, 589, 319

\end{thebibliography}

\end{document}